\documentclass[namedreferences]{solarphysics}
\usepackage[hyperref,optionalrh]{spr-sola-addons}
\usepackage{graphicx}
\usepackage{amssymb}            
\usepackage{color}                                        
\usepackage{breakurl}

\newcommand{\arcsec}{^{\prime\prime}}

\chardef\us=`\_

\begin{document}

\begin{article}
\begin{opening}
 
\title{Multiwavelength observations of a partial filament eruption on 13 June 2011}

\author[addressref={aff1,aff2},corref,email={zhangyj@pmo.ac.cn}]{\inits{Y.J.}\fnm{Yanjie}~\lnm{Zhang}\orcid{0000-0003-1979-9863}}
\author[addressref={aff1},corref,email={zhangqm@pmo.ac.cn}]{\inits{Q.M.}\fnm{Qingmin}~\lnm{Zhang}\orcid{0000-0003-4078-2265}}
\author[addressref={aff1,aff2},corref]{\inits{J. Dai}\fnm{Jun}~\lnm{Dai}\orcid{0000-0003-4787-5026}}
\author[addressref={aff1},corref]{\inits{D. Li}\fnm{Dong}~\lnm{Li}\orcid{0000-0002-4538-9350}}
\author[addressref={aff1},corref]{\inits{H.S. Ji}\fnm{Haisheng}~\lnm{Ji}\orcid{0000-0002-5898-2284}}

\address[id=aff1]{Key Laboratory of Dark Matter and Space Astronomy, Purple Mountain Observatory, CAS, Nanjing 210023, China}
\address[id=aff2]{School of Astronomy and Space Science, University of Science and Technology of China, Hefei 230026, China}

\runningauthor{Zhang et al.}
\runningtitle{A partial filament eruption on 13 June 2011}

\begin{abstract}
In this paper, we report the multiwavelength observations of the partial filament eruption associated with a C1.2 class flare in NOAA active region 11236 on 13 June 2011.
The event occurred at the eastern limb in the field of view (FOV) of Atmospheric Imaging Assembly (AIA) on board the Solar Dynamics Observatory (SDO) spacecraft
and was close to the disk center in the FOV of Extreme-UltraViolet Imager (EUVI) on board the behind Solar Terrestrial Relations Observatory (STEREO) spacecraft.
During eruption, the filament splits into two parts: the major part and runaway part. The major part flows along closed loops and experiences bifurcation at the loop top.
Some of the materials move forward and reach the remote footpoint, while others return back to the original footpoint.
The runaway part flows along open field lines, which is evidenced by a flare-related type III radio burst. 
The runaway part also undergoes bifurcation. The upper branch of escapes the corona and evolves into a jet-like narrow coronal mass ejection (CME) at a speed of $\sim$324 km s$^{-1}$, 
while the lower branch falls back to the solar surface. A schematic cartoon is proposed to explain the event and provides a new mechanism of partial filament eruptions.
\end{abstract}
\end{opening}

\section{Introduction}  \label{s-Intro}
Prominences are cool and dense plasmas in the solar atmosphere \citep{Lab2010,Mac2010,Par2014}. 
The dense materials are supported by the magnetic tension force of dips in sheared arcades or twisted flux ropes \citep{Pri1989,Aul1998,Ter2015,Gib2018}.
Prominences are usually observed in H$\alpha$, Ca {\sc ii} H, ultraviolet (UV), and extreme-ultraviolet (EUV) wavelengths above the solar limb \citep{Ber2010,Zha2012,Sch2014}.
They are also called filaments on the disk as a result of absorption of the background emissions \citep{Eng1998}. 
High-resolution observations reveal that filaments are composed of ultrafine dark threads \citep{Lin2005,Lin2008}.
A few hours to several days are needed to accumulate sufficient materials at filament channels along polarity inversion lines (PILs) \citep{Kar2005,Liu2012b,Yan2015,Li2018,Wang2019}.
Filaments with dextral (sinistral) chirality are predominant in the northern (southern) hemisphere \citep{Mar1998,OY2017}.

Prominences are rich in dynamics, such as counterstreaming flows \citep{Zhou2020}, upflows in dark plumes \citep{Ber2010,Hill2011}, 
rotation \citep{Su2012,Li2015,Hua2018}, small-amplitude oscillations \citep{Oka2007,Ning2009}, 
and large-amplitude oscillations \citep{Zha2012,Zha2017,Luna2018,Dai2021}. After being disturbed, prominences may become unstable and erupt outward \citep{Sch2013,Mc2015}, 
producing coronal jets \citep{Ster2015,Hong2016}, solar flares \citep{Zha2022b}, and normal/jet-like coronal mass ejections \citep[CMEs;][]{Nis2009,Hon2011,She2012,Pan2016b,Vou2017,Zha2022a}. 
There are basically three types of filament eruptions \citep{Gil2007}. The first type is failed or confined eruptions without CMEs \citep{Ji2003,Tor2005,Alex2006,Liu2009}. 
A filament rises to the maximum height and falls back to the solar surface as a result of strong compression of the overlying magnetic field.
The second type is full eruptions associated with CMEs during which a majority of filament material escapes into the interplanetary space \citep{Pats2010,Song2022}. 
The third type is partial eruptions during which filaments split into two parts with one part being fully erupted and the other staying behind \citep{Gil2000,Gil2001,Jos2014,Bi2015,Cheng2018}.
The degree to which a filament is expelled depends on where magnetic reconnection takes place.

\citet{Gib2006} performed a three-dimensional (3D) magnetohydrodynamics (MHD) numerical simulation of the partial eruption of a flux rope. 
After separating from the lower, surviving rope by multiple magnetic reconnection at current sheets, the upper rope escapes to drive a CME.
\citet{Tri2009} investigated six CMEs with partially-erupting prominences, which could well be explained by the partially-expelled-flux-rope model \citep{Gib2006,Gib2008}.
\citet{Liu2007} noticed the writhing motion and magnetic reconnection of a twisted, kink-unstable filament, which leads to the partial filament eruption.
The upper part above the reconnection point escapes to form a CME, while the lower part falls back.
For the first time, \citet{Liu2012a} put forward a scenario of double-decker filament, in which an upper flux rope is on top of a lower flux rope or a shear arcade before eruption (see their Fig. 12).
The magnetic system can be in equilibrium for a few hours or even days until the top flux rope erupts out 
and the low-lying branch remains stable \citep{Cheng2014,Kli2014,Awa2019,Zheng2019,Chen2021,Pan2021,Wei2021}.
Transfer of magnetic flux by internal reconnection plays a significant role for the upper branch to lose equilibrium and erupt smoothly \citep{Liu2012a}.
Interestingly, \citet{Chen2018} discovered a filament that split into three branches. The high-lying branch containing a flux rope erupted and produced a two-ribbon flare.
The low-lying two branches remained stable.
Besides, confined partial eruption of a double-decker filament has been reported, i.e., the upper branch experiences a failed eruption instead of a full eruption \citep{Zheng2019,Chen2021}.
Recently, high-resolution observations have given us a more detailed understanding of partial eruptions \citep{Liu2018,Poi2020,Dai2022}.
\citet{Mon2021} investigated a filament with bifurcated substructures, and the author found that the splitting of the filament was induced by the reconnection of the magnetic fields that enveloped the filament with surrounding loops, which confirmed the previous similar results \citep{Li2016,Xue2016}.
\citet{Jos2022} proposed a two reconnection process where one takes place below the filament, while another one occurs between the filament-carrying flux rope and the large-scale closed loops, which ultimately prevent the eruption.

Current observation shows that some filament eruptions are involved with the so-called ``two step'' process \citep{Byr2014,Gos2016,Cha2017}.
On this occasion, the filament exhibits a slow-rise phase and after reaching a certain height, it would gradually stop and remain stable for an hour \citep{Byr2014} to a day \citep{Gos2016}, until a full eruption develops, for which either kink instability or the torus instability is proposed to be the likeliest reason.
It usually results in two peaks in EUV and X-ray light curves, suggesting two periods of magnetic reconnection that takes place to change the topology of the system first and then facilitate the subsequent flux rope eruption. \citep{Woo2011,Su2012b}.
But the exact physical connection between these two stages is still unclear.

\citet{Zha2015} studied a partial filament eruption in NOAA active region (AR) 11283 on 8 September 2011. 
The $\mathsf{S}$-shaped filament was supported by a sheared arcade under a fan-spine dome related to a magnetic null point. During eruption, the filament split into two parts.
The major part reached the maximum height and returned to the solar surface in a bumpy way, which was associated with an M6.7 flare. 
The runaway part of the filament separated from the major part and escaped from the corona along open field lines, leading to a weak and narrow CME.
The presence of open field is also verified by the flare-related type III radio burst.
Narrow CMEs usually originate from blowout coronal jets extending to the field of view (FOV) of white-light (WL) coronagraphs \citep{Wang1998,Shen2012,Chan2017,Jos2018,Duan2022}.

Until now, observation of such kind of partial eruptions is still scarce, which is probably due to the accompanying CME is too weak to be detected \citep{Oka2021}.
In this paper, we carry out a detailed investigation of a similar event associated with a C1.2 class flare in NOAA AR 11236 (N17E91) on 13 June 2011.
We describe the data analysis in Section~\ref{s-Data}. The results are presented in Section~\ref{s-Res}.
A schematic cartoon is proposed to illustrate the partial eruption in Section~\ref{s-Dis}. Finally, a brief summary is given in Section~\ref{s-Sum}.

\section{Data Analysis} \label{s-Data}
\begin{figure}
\centerline{\includegraphics[width=0.9\textwidth,clip=]{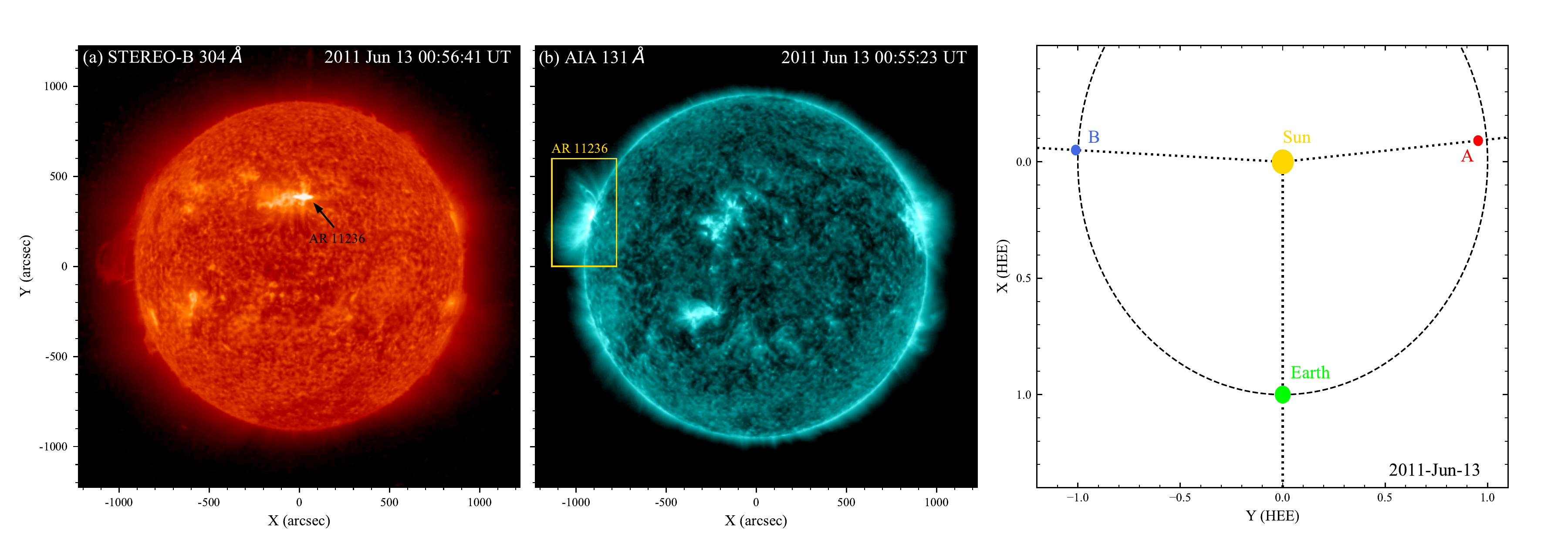}}
\caption{(a-b) The images of STEREO-B 304 {\AA} and AIA 131 {\AA} at 00:56 UT and 00:55 UT on 13 June 2013, respectively.
The location of AR 11236 is labelled in each panel.
The gold box shows the region where the intensity of AIA 131 {\AA} is calculated as presented in Figure~\ref{fig5}.
(c) Positions of the Earth, Ahead (A), and Behind (B) STEREO spacecrafts at 00:55 UT on 13 June 2011.}
\label{fig1}
\end{figure}

The left and middle panels of Figure~\ref{fig1} show the images of the behind Solar Terrestrial Relations Observatory \citep[STEREO;][]{Kai2008} spacecraft and the Atmospheric Imaging Assembly \citep[AIA;][]{Lem2012} on board the Solar Dynamics Observatory \citep[SDO;][]{Pes2012} spacecraft on 13 June 2011 respectively, where the AR 11236 is labelled.
The prominence eruption and C1.2 flare occurred at the eastern limb in the field of FOV of SDO/AIA, which 
takes full-disk EUV images out to 1.3\,$R_{\odot}$ with a cadence of 12\,s and a spatial resolution of 1.2$\arcsec$.
The AIA level\_1 data were calibrated using the standard Solar Software (SSW) program \textsf{aia\_prep.pro}.
The eruption was also observed by the ground-based telescope of Big Bear Solar Observatory (BBSO) in H$\alpha$ line center (6562.8 {\AA}) 
with a cadence of 60\,s and a spatial resolution of 2.08$\arcsec$.
The corresponding narrow CME was observed by the Large Angle Spectroscopic Coronagraph \citep[LASCO;][]{Bru1995} on board SOHO spacecraft
and recorded by the CDAW CME catalogue\footnote{http://cdaw.gsfc.nasa.gov/CME\_list}.
The LASCO-C2 coronagraph has a FOV of 2$-$6\,$R_{\odot}$.

Figure~\ref{fig1}(c) shows positions of the Earth, Ahead (A), and Behind (B) STEREO spacecraft on 13 June 2011.
The separation angle of STEREO-B with the Earth was $\sim$92$^{\circ}$. Therefore, the filament in AR 11236 was close to the central meridian in the FOV of 
Extreme-UltraViolet Imager \citep[EUVI;][]{Wue2004} of the Sun Earth Connection Coronal and Heliospheric Investigation \citep[SECCHI;][]{How2008} on board STEREO-B.
AIA and EUVI provided an edge-on and a head-on view of the filament eruption, respectively.
EUVI takes full-disk images out to 1.7\,$R_{\odot}$ with a spatial resolution of 3.2$\arcsec$ in 171, 195, 284, and 304 {\AA}. The 195 and 304 {\AA} images have time cadences of 2.5 minutes.
Soft X-ray (SXR) light curves of the C1.2 class flare in 0.5$-$4 {\AA} and 1$-$8 {\AA} were recorded by the Geostationary Operational Environmental Satellite (GOES) spacecraft.
Hard X-ray (HXR) light curves at 5$-$12 and 12$-$27\,keV were observed by the Gamma-ray Burst Monitor \citep[GBM;][]{Mee2009} on board the \textit{Fermi} spacecraft. 
A type III radio burst related to the flare was observed by the S/WAVES instrument \citep{Bou2008} on board STEREO-B. 

\section{Results} \label{s-Res}
\begin{figure} 
\centerline{\includegraphics[width=1\textwidth,clip=]{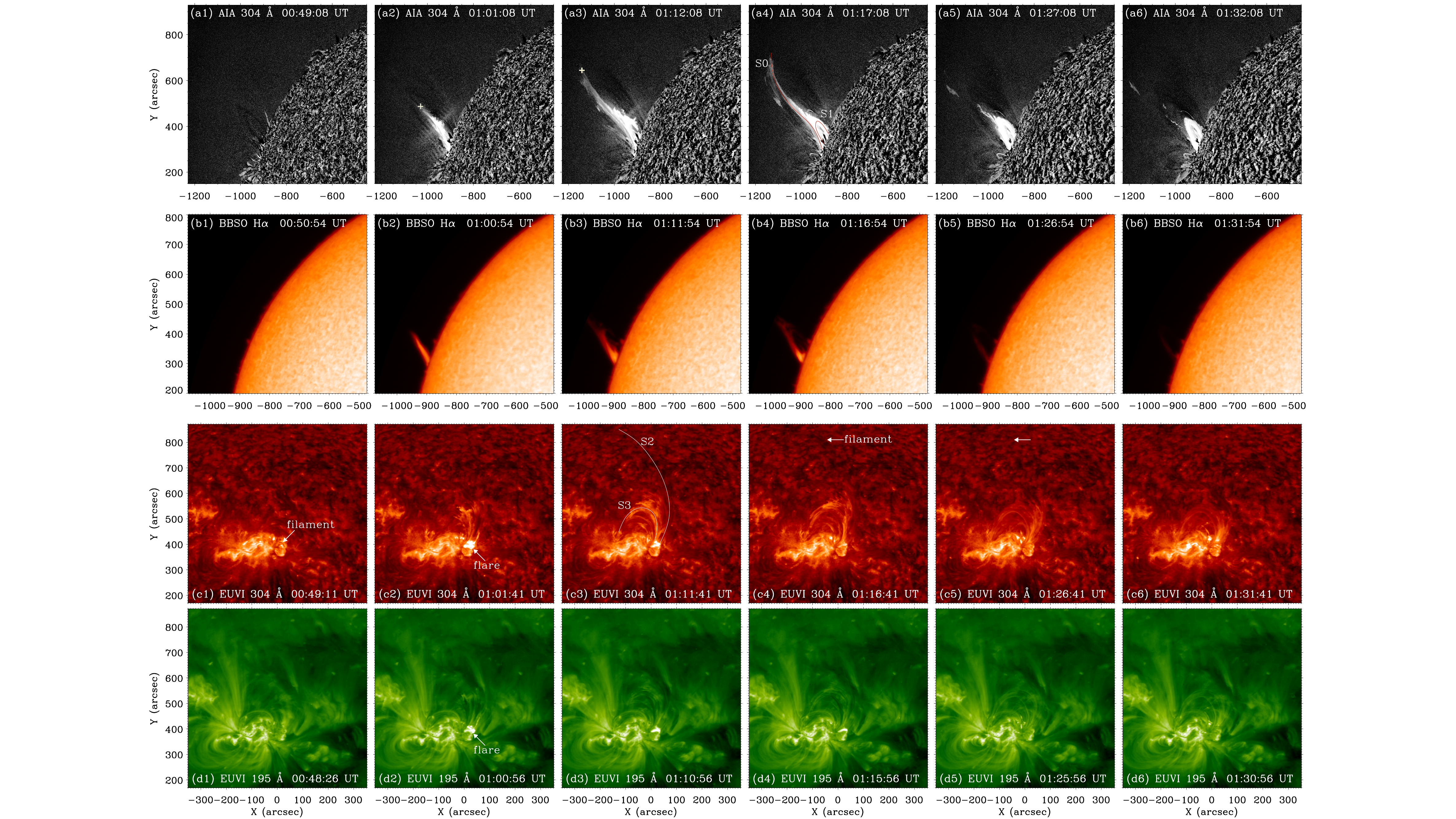}}
\caption{From top to bottom: EUV images of the prominence/filament eruption during 00:49$-$01:32 UT
observed by AIA 304 {\AA} (base-difference images), BBSO H$\alpha$ line center, EUVI-B 304 {\AA}, and EUVI-B 171 {\AA}, respectively.
Four artificial slices are selected to investigate the kinematics of filament: S0 and S1 in panel (a4), S2 and S3 in panel (c3). The white arrows point to the filament and C1.2 class flare.
Arrows in panel (c4) and (c5) are combined to highlight the escaping part of the filament which was too diffuse to be observed in static images but visible in the animation attached to the figure,
which is available in the Electronic Supplementary Material (\textsf{Fig2.mp4}).
The plus symbols in panels (a2) and (a3) track the leading edge of the erupting filament, which is plotted in Figure~\ref{fig4}.}
\label{fig2}
\end{figure}

In Figure~\ref{fig2}, the top two rows show the evolution of prominence observed by AIA in 304 {\AA} (base-difference images) and BBSO in H$\alpha$ line center 
(see also online animation \textsf{Fig2.mp4}). The prominence starts to erupt from AR 11236 at $\sim$00:50 UT.
During eruption, most of the materials, i.e., the major part, rises and flows along closed loops, reaching the opposite (northwest) footpoint (panels (a3-a6)).
A minority of materials, i.e., the runaway part, flows out of the corona northeastward along a curved trajectory (S0). That is to say, the filament is divided into the major part and runaway part.
In H$\alpha$ wavelength, the eruption of major part is evident as well, while the runaway part is undistinguishable.

The bottom two rows of Figure~\ref{fig2} show the evolution of filament observed by EUVI-B in 304 and 195 {\AA}. The eruption of filament generates a flare at the western footpoint (panels (c2, d2)). 
The major part flows along closed loops and terminates at the eastern footpoint (panel (c3)), 
while the runaway part propagates along S2 and shows absorption compared with the background (panel (c4)). 
Hence, the two parts are clearly demonstrated in EUVI-B as well.

\begin{figure}
\centerline{\includegraphics[width=0.8\textwidth,clip=]{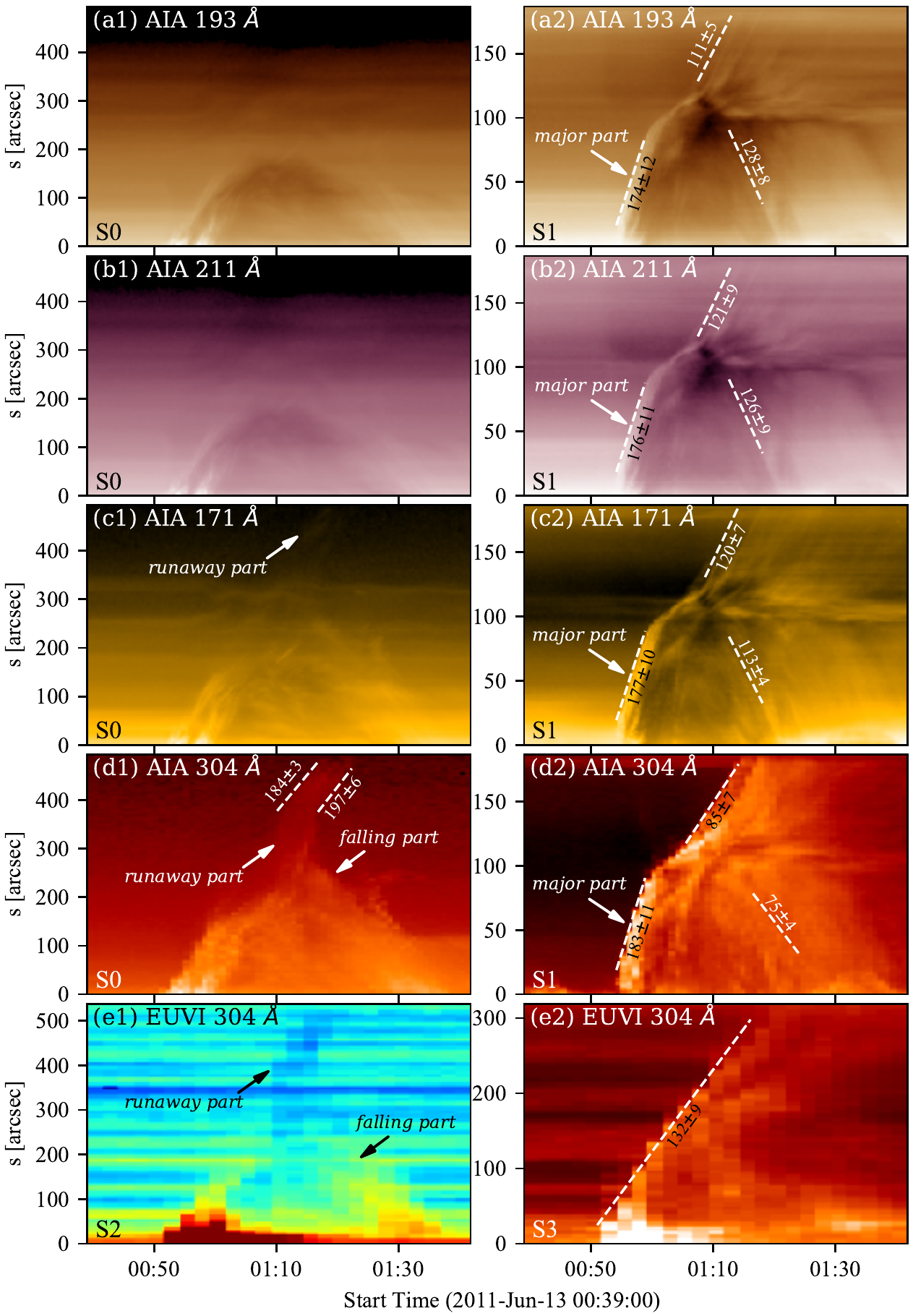}}
\caption{(a1)-(d1) Time-slice diagrams of S0 in AIA 193, 211, 171, and 304 {\AA}. (e1) Time-slice diagram of S2 in EUVI-B 304 {\AA}.
The arrows point to the runaway part and falling part. The velocities of runaway part and falling part are labeled, whose units is km s$^{-1}$.
(a2)-(d2) Time-slice diagrams of S1 in AIA 193, 211, 171, and 304 {\AA}. (e2) Time-slice diagram of S3 in EUVI-B 304 {\AA}.
The apparent velocities of major part are labeled.}
\label{fig3}
\end{figure}

To investigate the kinematics of the two parts, we select four artificial slices in Figure~\ref{fig2}: S0 and S1 in panel (a4), S2 and S3 in panel (c3). S1 and S3 are along closed field lines.
All the slices are 10$\arcsec$ in width.
Time-distance diagrams of S1 in AIA 193, 211, 171, and 304 {\AA} are displayed in Figure~\ref{fig3}(a2-d2).
Time-distance diagram of S3 in EUVI-B 304 {\AA} is displayed in Figure~\ref{fig3}(e2). 
The major part rises rapidly along S1 to the loop top at speeds of 178$\pm$11 km s$^{-1}$. 
Afterwards, part of the material moves forward along the same loops at speeds of 109$\pm$7 km s$^{-1}$ and stops at the opposite footpoint around 01:15 UT.
However, the residual plasma returns back to the original footpoint continuously at speeds of 110$\pm$6 km s$^{-1}$, suggesting that the major part undergoes bifurcation at the loop top.

S0 and S2 are along open field lines. Time-distance diagrams of S0 in AIA 193, 211, 171, and 304 {\AA} are displayed in Figure~\ref{fig3}(a1-d1).
Time-distance diagram of S2 in EUVI-B 304 {\AA} is displayed in Figure~\ref{fig3}(e1).
The filament eruption commenced at $\sim$00:50 UT, while the runaway part began to escape the corona at $\sim$01:09 UT with an apparent speed of 191$\pm$5 km s$^{-1}$ (panel (d1)).
The runaway part is most evident in 304 {\AA} and slightly visible in 171 {\AA}. 
In 193 and 211 {\AA}, the runaway part is almost absent, indicating that this part is not heated to higher temperatures.
Besides, some of the plasmas fall back to the solar surface at $\sim$01:30 UT (panels (d1, e1)).

\begin{figure}
\centerline{\includegraphics[width=0.9\textwidth,clip=]{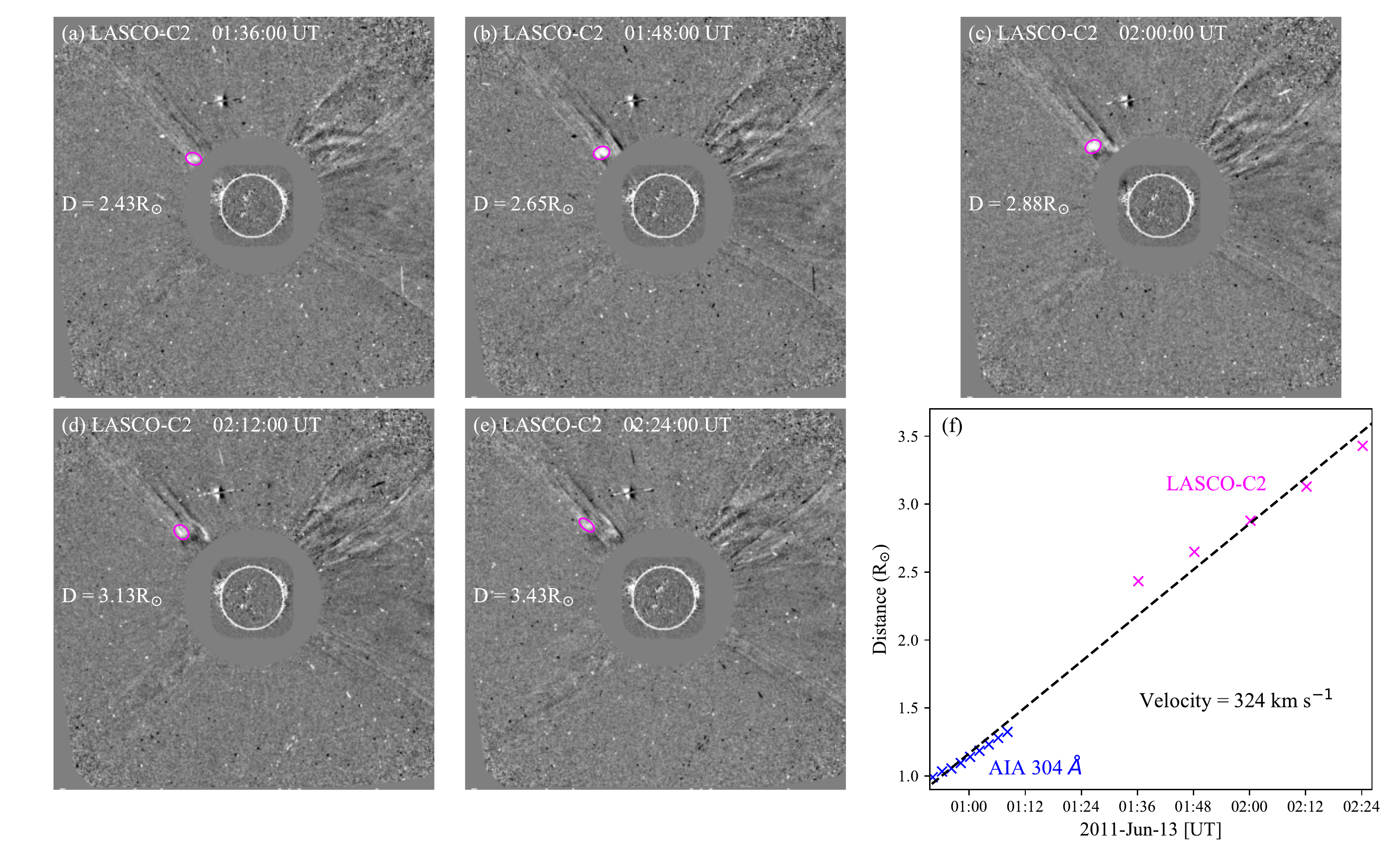}}
\caption{(a)-(e) Running-difference images of the jet-like CME observed by LASCO-C2 during 01:36$-$02:24 UT.
The leading fronts are encircled with magenta ellipses and their heliocentric distances are labeled.
(f) Height-time plot of the filament/CME leading edges with blue/magenta cross symbols. The results of linear fit is overploted with a dashed line and the linear speed (324 km s$^{-1}$) is labeled.}
\label{fig4}
\end{figure}

As soon as the runaway part enters the FOV of LASCO-C2, it produces a faint jet-like CME with a central position angle of 45$^{\circ}$ and an angular width of 23$^{\circ}$, respectively.
Figure~\ref{fig4}(a-e) show five running-difference images of LASCO-C2 during 01:36$-$02:24 UT, 
in which the leading edges of CME are encircled by magenta ellipses and their heliocentric distances are labeled.
We also track the heliocentric distances of the leading edge of erupting prominence in AIA 304 {\AA} during 00:52 UT $\sim$ 01:12 UT as depicted in Figure~\ref{fig2}(a2) and (a3).
Figure~\ref{fig4}(f) shows the height-time plot of these figures with a linear speed of 324 km s$^{-1}$ (dashed line).
The consistency between the two kinds of figures indicates that the jet-like CME does originate from the erupting filament.
While the speed recorded on the CDAW CME catalogue is 427 km s$^{-1}$. The reason for the difference between the two values is that the CME underwent an accelerated process as illustrated in the CDAW CME catalogue, and only the initial process with the visible leading edge is linearly fitted by us.
It should be emphasized that the runaway part of filament along the trajectory of S2 become too faint to be detected as a CME in the FOV of STEREO-COR1 coronagraph.

\begin{figure}
\centerline{\includegraphics[width=0.9\textwidth,clip=]{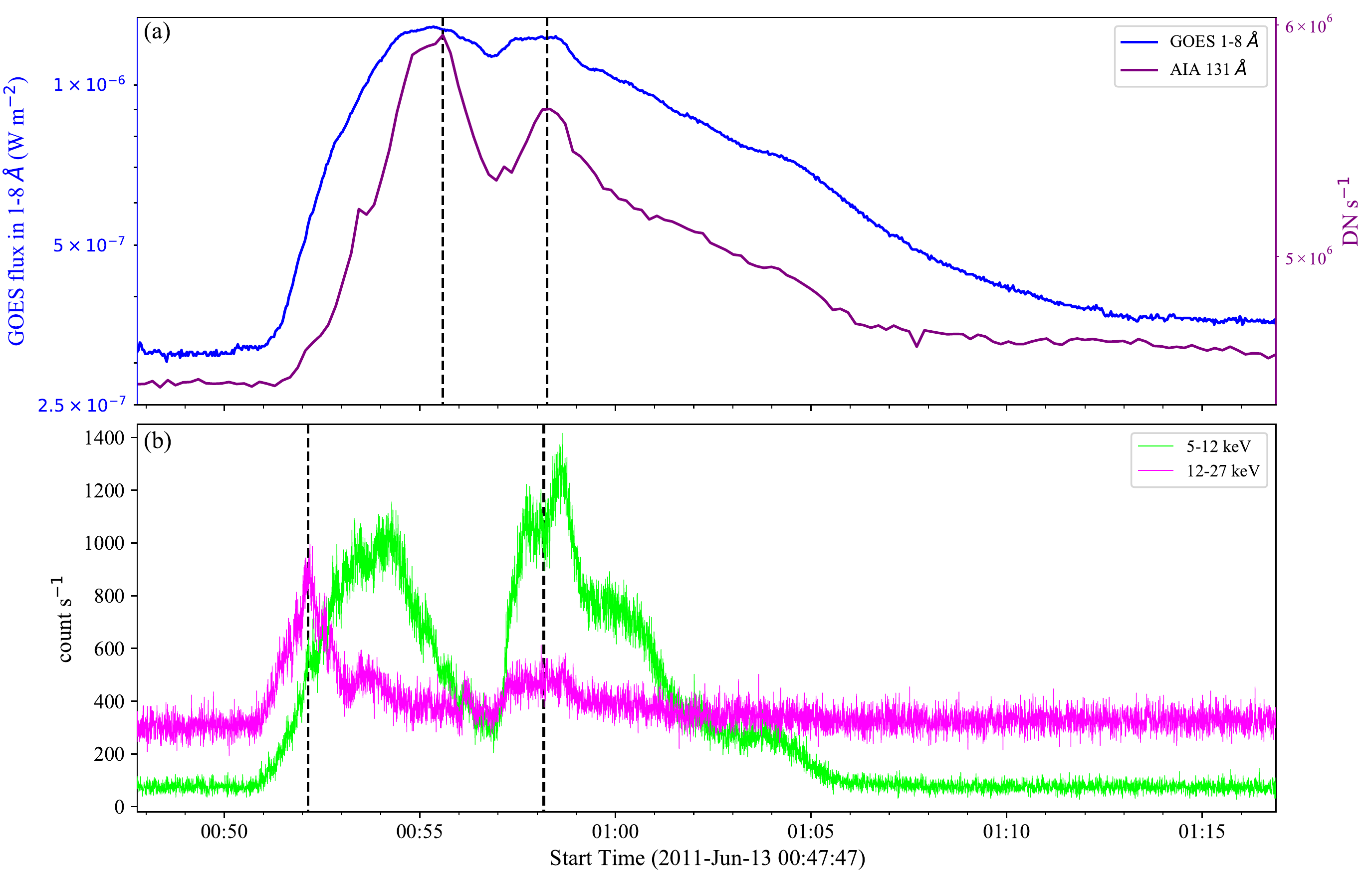}}
\caption{(a) SXR 1$-$8 {\AA} light curve (blue line) and EUV 131 {\AA} light curve (purple line) of the C1.2 flare.
The vertical dashed lines denote the two peaks at 00:55:30 UT and 00:58:15 UT.
(b) HXR light curves of the flare at 5$-$12 keV (green line) and 12$-$27 keV (magenta line).
The vertical dashed lines denote the two peaks (00:52:10 UT and 00:58:10 UT) at 12$-$27 keV.}
\label{fig5}
\end{figure}

In Figure~\ref{fig5}, the top panel shows the SXR light curve of the associated C1.2 flare in 1$-$8 {\AA} (blue line). 
The flux starts to increase at $\sim$00:50 UT and reaches the first peak at $\sim$00:55:30 UT, 
which is followed by a second peak at $\sim$00:58:15 UT and a long decay phase until $\sim$01:15 UT. Hence, the lifetime of the flare is $\sim$25 minutes.
The light curve in AIA 131 {\AA}, integrated over the box in Figure~\ref{fig1}(b), is plotted with a purple line, which is characterized by the same two peaks as in 1$-$8 {\AA}. 
The consistency between the light curves in SXR and EUV suggests that both of the two peaks are from the same flare with two episodes of energy release, rather than two separate flares.
Because the 131 {\AA} channel light curve, which indicates a significant high temperature plasma contribution (T$\approx$10 MK), tracks the GOES 1-8 {\AA} flux best of all the AIA passband \citep{Fle2013,Tia2015}.
HXR light curves of the flare at 5$-$12 keV (green line) and 12$-$27 keV (magenta line) are depicted in Figure~\ref{fig5}(b).
Two major peaks can also be identified in these energy bands. The two peaks (00:52:10 UT and 00:58:10 UT) at 12$-$27 keV are denoted by two vertical dashed lines.
Since the flare ribbons or kernels were blocked by the eastern limb, HXR emissions of higher energy bands ($>$27 keV) originating from the footpoint sources were negligible \citep{Liu2006}.
Meanwhile, the flare was radio quiet at frequencies of $\geq$1 GHz recorded by the Nobeyama Radio Polarimeters\footnote{http://solar.nro.nao.ac.jp/norp/html/daily/}.

\begin{figure}
\centerline{\includegraphics[width=0.6\textwidth,clip=]{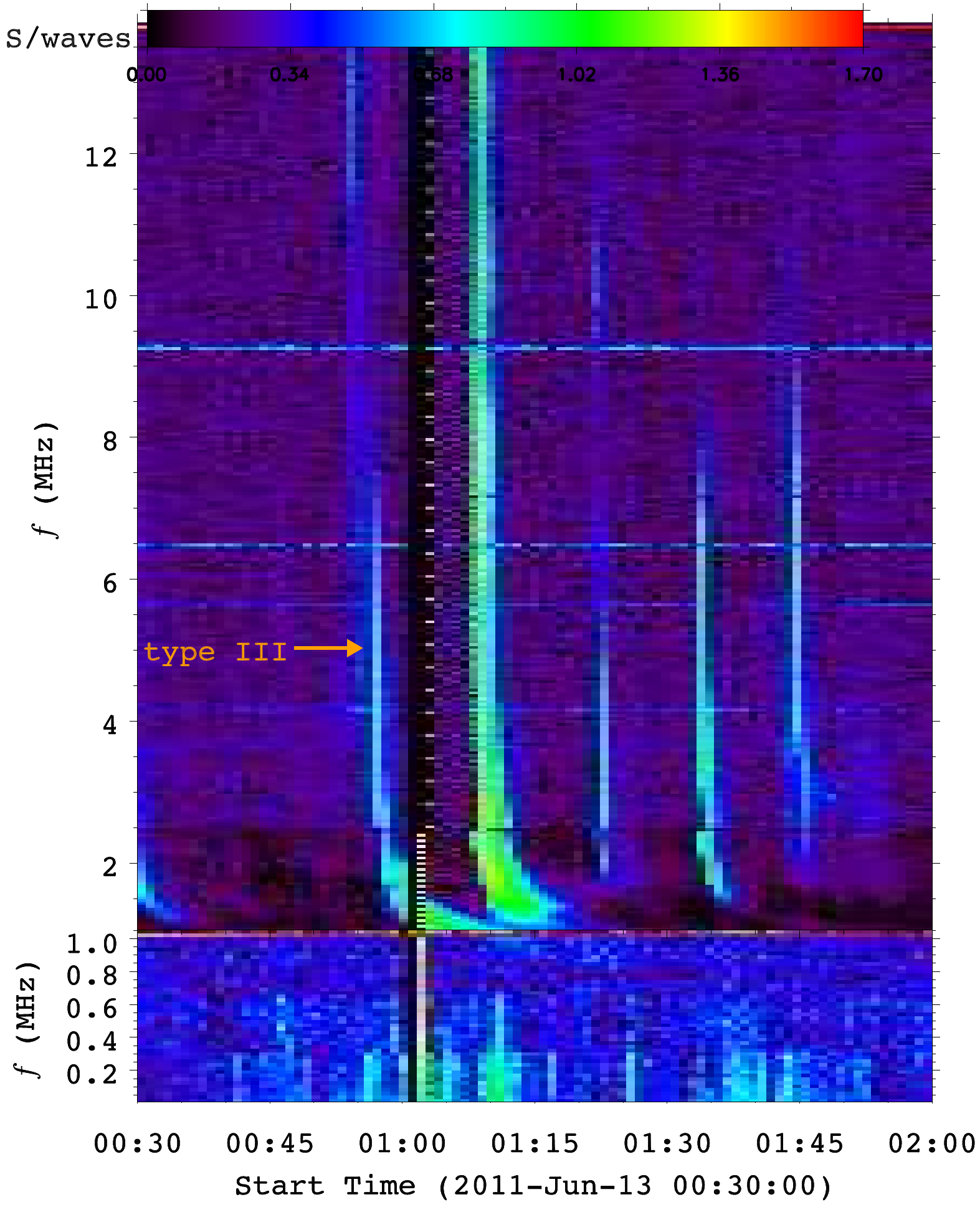}}
\caption{Radio dynamic spectra observed by S/WAVES on board STEREO-B.
The arrow points to the flare-related type III radio burst, which undergoes fast frequency drift during 00:56$-$01:14 UT.}
\label{fig6}
\end{figure}

A flare-related type III radio burst was detected by S/WAVES on board STEREO-B as shown in Figure~\ref{fig6}.
The frequency drifted rapidly from $\sim$10 MHz to $\sim$1 MHz during 00:56$-$01:14 UT, which is well consistent with the second peak in HXR.
Considering that type III radio bursts are created by plasma emissions of flare-accelerated non-thermal electron beams propagating outward along open field,
the type III radio burst around 00:58 UT confirms the existence of open magnetic field.
Combining Figure~\ref{fig4} and Figure~\ref{fig6}, it is concluded that the open field provides a tunnel not only for jet-like CME, 
but also for nonthermal electrons \citep{Kru2011,Mas2013,Zha2015,Wyp2018}.
 
\section{Plausible physical explanation of the partial eruption} \label{s-Dis}
As mentioned in Section~\ref{s-Intro}, partial filament eruptions are frequently observed. However, open magnetic field is not involved in previous models \citep{Gil2000,Gil2001,Gib2006,Liu2012a}. 
Moreover, internal magnetic reconnection plays an important role in the vertical splitting of a filament into two parts or branches \citep[e.g.,][]{Liu2007,Chen2018,Cheng2018}. 
However, the current event on 13 June 2011 and the event on 8 September 2011 \citep{Zha2015} are different from these models.
On one hand, open field is present and serves as a tunnel for the runaway part of filament to escape the corona and evolves into a narrow CME.
Type III radio bursts are produced when flare-accelerated nonthermal electrons propagate along open field lines. On the other hand, the splitting of a filament is not caused by internal reconnection.

\begin{figure}
\centerline{\includegraphics[width=0.9\textwidth,clip=]{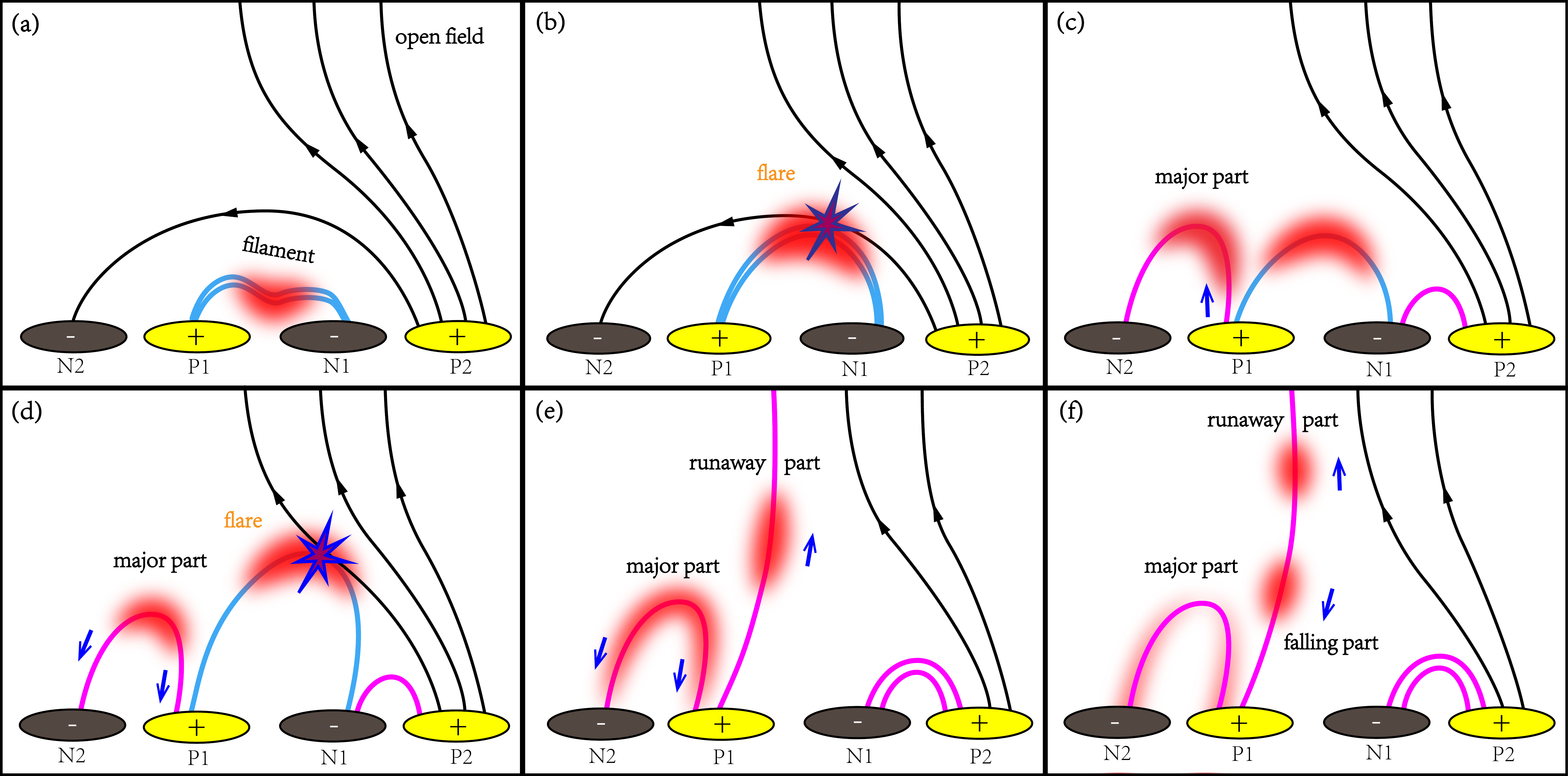}}
\caption{A schematic cartoon to illustrate the partial eruption on 13 June 2011. The filament (red clouds) is supported by a magnetic sheared arcade (blue lines).
Successive magnetic reconnections take place between the ascending filament and the closed and open field lines, producing the major part and runaway part, respectively.
Both of them undergo bifurcation. The short blue arrows indicate the directions of mass flows. The two-stage reconnections correspond to the two peaks in the light curves of related flare.}
\label{fig7}
\end{figure}

In Figure~\ref{fig7}, we draw a schematic cartoon to explain the event. Before eruption, the filament is supported by a magnetic sheared arcade (panel (a)). 
After being disturbed, such as flux emergence \citep{Chen2000} or shearing motion \citep{Zha2015} at the photosphere, the filament starts to rise.
As soon as the filament encounters and collides with the overlying closed field of opposite magnetic polarity, magnetic reconnection takes place and a flare is produced (panel (b)),
which corresponds to the first peak in Figure~\ref{fig5}. Magnetic reconnection between an ascending filament and high-lying coronal loops has been reported before \citep{Li2016,Xue2016,Sri2021}.
The nonthermal electrons propagate downward to the chromosphere, where HXR emissions are generated \citep{Bro1971}.
Since nonthermal electrons are restricted to the closed loops at this stage, a type III radio burst is absent. 
The major part of filament flows along the reconnected closed loop, while the remaining filament continues to rise (panel (c)). At the loop top, the major part undergoes bifurcation (panel (d)). 
Some of the plasmas continue to move forward and reach the remote footpoint (N2), while the residual plasmas fall back to the starting footpoint (P1) (see Figure~\ref{fig3}(a2-e2)).
Once the remaining filament collides with open field lines, magnetic reconnection takes place (panel (d)), which corresponds to the second peak in Figure~\ref{fig5}.
A type III radio burst is created as the nonthermal electrons escape the corona along open field lines (Figure~\ref{fig6}). 
Meanwhile, the runaway part of filament propagates along open field lines (panel (e)). 
The interchange reconnection is similar to jet models where newly emerging flux reconnects with previously existing open field \citep{Shi1992,Mor2008,Arc2013}.
Interestingly, the runaway part also undergoes bifurcation (panel (f)).
Those propagating upward forms a jet-like CME (Figure~\ref{fig4}), while those flowing downward (falling part) return to the solar surface (Figure~\ref{fig3}(a1-e1)).
In brief, the splitting of filament into a major part and a runaway part is not realized by internal reconnection, but by two-step reconnections with the closed field and open field, respectively.
Both of the parts experience bifurcation during their evolution, showing a more complex behavior than before.
It is noted that although the filaments are supported by sheared arcades in this event and the previous event on 8 September 2011, 
the two-step magnetic reconnections could also take place sequentially between a filament-carrying flux rope and the overlying closed and open fields.
In this circumstance, rotation is expected in the major part and runaway part \citep{Zha2014,Li2015,Xue2016}.
Besides, there are not necessarily two separate peaks in SXR light curves of the related flare if the two-step reconnections are fast enough.

This model has important implications. Firstly, it shows a new picture and represents a new category of partial eruptions, which is completely different from previous models. 
Traditionally, the upper part or branch erupts to create a CME, while the lower part or branch leaves behind.
In our model, the filament materials with higher latitudes reconnect with closed loops and develop into the major part, which fails to escape.
Subsequently, the filament materials with lower latitudes reconnect with open field lines and develop into a CME successfully.
In a sense, the first-step reconnection removes constraint from the overlying closed field, which facilitates the eruption of lower filament materials.
Secondly, a jet-like narrow CME observed by WL coronagraphs and a ``failed'' eruption observed on the solar disk may stem from the same partial eruption. 
Therefore, stereoscopic observations are required to have a full view of filament eruptions, 
combining the capabilities of SDO/AIA, SOHO/LASCO, STEREO, Metis \citep{Ant2020} on board Solar Orbiter \citep{Mul2020}, 
and the Wide-Field Imager for Solar Probe Plus \citep[WISPR;][]{Vour2016} on board the Parker Solar Probe \citep[PSP;][]{Fox2016} in the future.

Of course, the current investigation has limitation. Since the prominence eruption and flare took place at the eastern limb, 
reliable photospheric magnetograms and nonlinear force-free field (NLFFF) modeling are unavailable. Therefore, the nonpotential magnetic configuration before flare is still unclear.
A statistical analysis is under way to find more cases and investigate their origin, which will be the topic of our next paper.

\section{Summary} \label{s-Sum}
In this paper, we report the multiwavelength observations of the partial filament eruption associated with a C1.2 class flare in AR 11236 on 13 June 2011.
The event occurs at the eastern limb in the FOV of SDO/AIA and is close to the disk center in the FOV of EUVI-B.
Main results are summarized as follows. 
\begin{enumerate}
\item{During eruption, the filament splits into two parts: the major part and runaway part. The major part flows along closed loops and experiences bifurcation at the loop top.
Some of the materials move forward and reach the opposite footpoint, while others return back to the original footpoint.}
\item{The runaway part flows along open field lines, which is suggested by a flare-related type III radio burst.
The upper branch escapes the corona and evolves into a jet-like narrow CME at a speed of $\sim$324 km s$^{-1}$, while the lower branch falls back to the solar surface.}
\end{enumerate}
Unlike the well-known mechanisms of partial filament eruption \citep{Gib2006,Liu2012a}, This prominence in our paper got separated during the outward ejection, rather than just at the beginning of the eruption.
In our proposed model, the key point is the reconnection of the magnetic structure of prominence with the surrounding open magnetic fields, which leads to the escape of plasma.
While the reconnection with overlying closed magnetic fields (the falling back of the major part) as well as the gravity (the bifurcation of the runaway part) prevents the rest from escaping, which makes a difference with the interchange jet models \citep{Shi1992}.
However, there exist other ways to cause a partial eruption in this scenario, such as the asymmetry of the background magnetic fields with respect to the position of the filament as proposed by \citet{Liu2009,Zha2015}.
Therefore, in the future, partial filament eruptions accompanied by jet-like CME should be paid more attention to replenish this model.

\begin{acks}
The authors are grateful to the colleagues in Purple Mountain Observatory for their constructive suggestions and comments.
SDO is a mission of NASA\rq{}s Living With a Star Program. AIA and HMI data are courtesy of the NASA/SDO science teams.
STEREO/SECCHI data are provided by a consortium of US, UK, Germany, Belgium, and France.
This work is supported by the National Key R\&D Program of China 2021 YFA1600500 (2021YFA1600502) and NSFC grants (No. 11790302, 11790300, 11973092).
\end{acks}


\bibliographystyle{spr-mp-sola}
\bibliography{ms}

\begin{thebibliography}{110}
\ifx\bisbn     \undefined \def\bisbn  #1{ISBN #1}\fi
\ifx\binits    \undefined \def\binits#1{#1}\fi
\ifx\bauthor   \undefined \def\bauthor#1{#1}\fi
\ifx\batitle   \undefined \def\batitle#1{#1}\fi
\ifx\bjtitle   \undefined \def\bjtitle#1{\textit{#1}}\fi
\ifx\bvolume   \undefined \def\bvolume#1{\textbf{#1}}\fi
\ifx\byear     \undefined \def\byear#1{#1}\fi
\ifx\bissue    \undefined \def\bissue#1{#1}\fi
\ifx\bfpage    \undefined \def\bfpage#1{#1}\fi
\ifx\blpage    \undefined \def\blpage #1{#1}\fi
\ifx\burl      \undefined \def\burl#1{#1}\fi
\ifx\href      \undefined \def\href#1#2{#2}\fi
\ifx\betal     \undefined \def\betal{et al.}\fi
\ifx\bctitle   \undefined \def\bctitle#1{#1}\fi
\ifx\beditor   \undefined \def\beditor#1{#1}\fi
\ifx\bbtitle   \undefined \def\bbtitle#1{\textit{#1}}\fi
\ifx\bedition  \undefined \def\bedition#1{#1}\fi
\ifx\bseriesno \undefined \def\bseriesno#1{\textbf{#1}}\fi
\ifx\blocation \undefined \def\blocation#1{#1}\fi
\ifx\bsertitle \undefined \def\bsertitle#1{\textit{#1}}\fi
\ifx\bsnm      \undefined \def\bsnm#1{#1}\fi
\ifx\bsuffix   \undefined \def\bsuffix#1{#1}\fi
\ifx\bparticle \undefined \def\bparticle#1{#1}\fi
\ifx\barticle  \undefined \def\barticle#1{}\fi
\ifx\binstitute  \undefined \def\binstitute#1{#1}\fi
\ifx\bpublisher  \undefined \def\bpublisher#1{#1}\fi
\ifx\doiurl    \undefined \def\doiurl#1{\href{#1}{DOI}}\fi
\makeatletter
\def\safeHref#1#2#3{\in@{http}{#2}\ifin@\href{#2}{#3}\else\href{#1#2}{#3}\fi}
\makeatother
\ifx\adsurl    \undefined
  \def\adsurl#1{\safeHref{https://ui.adsabs.harvard.edu/abs/}{#1}{ADS}}\fi
\ifx\arxivurl  \undefined
  \def\arxivurl#1{\safeHref{http://arxiv.org/abs/}{#1}{arXiv}}\fi
\ifx\botherref \undefined \def\botherref#1{}\fi
\ifx\url       \undefined \def\url#1{#1}\fi
\ifx\bchapter  \undefined \def\bchapter#1{}\fi
\ifx\bbook     \undefined \def\bbook#1{}\fi
\ifx\bcomment  \undefined \def\bcomment#1{#1}\fi
\ifx\oauthor   \undefined \def\oauthor#1{#1}\fi
\ifx\citeauthoryear \undefined\def \citeauthoryear#1{#1}\fi
\def\endbibitem {}
\ifx\bconflocation  \undefined \def\bconflocation#1{#1} \fi

\bibitem[\protect\citeauthoryear{{Alexander}, {Liu}, and
  {Gilbert}}{2006}]{Alex2006}
\begin{barticle}
\bauthor{\bsnm{{Alexander}}, \binits{D.}},
\bauthor{\bsnm{{Liu}}, \binits{R.}},
\bauthor{\bsnm{{Gilbert}}, \binits{H.R.}}:
\byear{2006},
\batitle{{Hard X-Ray Production in a Failed Filament Eruption}}.
\bjtitle{\apj}
\bvolume{653},
\bfpage{719}.
\doiurl{https://doi.org/10.1086/508137}.
\adsurl{2006ApJ...653..719A}.
\end{barticle}
\endbibitem

\bibitem[\protect\citeauthoryear{{Antonucci} et~al.}{2020}]{Ant2020}
\begin{barticle}
\bauthor{\bsnm{{Antonucci}}, \binits{E.}},
\bauthor{\bsnm{{Romoli}}, \binits{M.}},
\bauthor{\bsnm{{Andretta}}, \binits{V.}},
\bauthor{\bsnm{{Fineschi}}, \binits{S.}},
\bauthor{\bsnm{{Heinzel}}, \binits{P.}},
\bauthor{\bsnm{{Moses}}, \binits{J.D.}},
\bauthor{\bsnm{{Naletto}}, \binits{G.}},
\bauthor{\bsnm{{Nicolini}}, \binits{G.}},
\bauthor{\bsnm{{Spadaro}}, \binits{D.}},
\bauthor{\bsnm{{Teriaca}}, \binits{L.}},
\bauthor{\bsnm{{Berlicki}}, \binits{A.}},
\bauthor{\bsnm{{Capobianco}}, \binits{G.}},
\bauthor{\bsnm{{Crescenzio}}, \binits{G.}},
\bauthor{\bsnm{{Da Deppo}}, \binits{V.}},
\bauthor{\bsnm{{Focardi}}, \binits{M.}},
\bauthor{\bsnm{{Frassetto}}, \binits{F.}},
\bauthor{\bsnm{{Heerlein}}, \binits{K.}},
\bauthor{\bsnm{{Landini}}, \binits{F.}},
\bauthor{\bsnm{{Magli}}, \binits{E.}},
\bauthor{\bsnm{{Marco Malvezzi}}, \binits{A.}},
\bauthor{\bsnm{{Massone}}, \binits{G.}},
\bauthor{\bsnm{{Melich}}, \binits{R.}},
\bauthor{\bsnm{{Nicolosi}}, \binits{P.}},
\bauthor{\bsnm{{Noci}}, \binits{G.}},
\bauthor{\bsnm{{Pancrazzi}}, \binits{M.}},
\bauthor{\bsnm{{Pelizzo}}, \binits{M.G.}},
\bauthor{\bsnm{{Poletto}}, \binits{L.}},
\bauthor{\bsnm{{Sasso}}, \binits{C.}},
\bauthor{\bsnm{{Sch{\"u}hle}}, \binits{U.}},
\bauthor{\bsnm{{Solanki}}, \binits{S.K.}},
\bauthor{\bsnm{{Strachan}}, \binits{L.}},
\bauthor{\bsnm{{Susino}}, \binits{R.}},
\bauthor{\bsnm{{Tondello}}, \binits{G.}},
\bauthor{\bsnm{{Uslenghi}}, \binits{M.}},
\bauthor{\bsnm{{Woch}}, \binits{J.}},
\bauthor{\bsnm{{Abbo}}, \binits{L.}},
\bauthor{\bsnm{{Bemporad}}, \binits{A.}},
\bauthor{\bsnm{{Casti}}, \binits{M.}},
\bauthor{\bsnm{{Dolei}}, \binits{S.}},
\bauthor{\bsnm{{Grimani}}, \binits{C.}},
\bauthor{\bsnm{{Messerotti}}, \binits{M.}},
\bauthor{\bsnm{{Ricci}}, \binits{M.}},
\bauthor{\bsnm{{Straus}}, \binits{T.}},
\bauthor{\bsnm{{Telloni}}, \binits{D.}},
\bauthor{\bsnm{{Zuppella}}, \binits{P.}},
\bauthor{\bsnm{{Auch{\`e}re}}, \binits{F.}},
\bauthor{\bsnm{{Bruno}}, \binits{R.}},
\bauthor{\bsnm{{Ciaravella}}, \binits{A.}},
\bauthor{\bsnm{{Corso}}, \binits{A.J.}},
\bauthor{\bsnm{{Alvarez Copano}}, \binits{M.}},
\bauthor{\bsnm{{Aznar Cuadrado}}, \binits{R.}},
\bauthor{\bsnm{{D'Amicis}}, \binits{R.}},
\bauthor{\bsnm{{Enge}}, \binits{R.}},
\bauthor{\bsnm{{Gravina}}, \binits{A.}},
\bauthor{\bsnm{{Jej{\v{c}}i{\v{c}}}}, \binits{S.}},
\bauthor{\bsnm{{Lamy}}, \binits{P.}},
\bauthor{\bsnm{{Lanzafame}}, \binits{A.}},
\bauthor{\bsnm{{Meierdierks}}, \binits{T.}},
\bauthor{\bsnm{{Papagiannaki}}, \binits{I.}},
\bauthor{\bsnm{{Peter}}, \binits{H.}},
\bauthor{\bsnm{{Fernandez Rico}}, \binits{G.}},
\bauthor{\bsnm{{Giday Sertsu}}, \binits{M.}},
\bauthor{\bsnm{{Staub}}, \binits{J.}},
\bauthor{\bsnm{{Tsinganos}}, \binits{K.}},
\bauthor{\bsnm{{Velli}}, \binits{M.}},
\bauthor{\bsnm{{Ventura}}, \binits{R.}},
\bauthor{\bsnm{{Verroi}}, \binits{E.}},
\bauthor{\bsnm{{Vial}}, \binits{J.-C.}},
\bauthor{\bsnm{{Vives}}, \binits{S.}},
\bauthor{\bsnm{{Volpicelli}}, \binits{A.}},
\bauthor{\bsnm{{Werner}}, \binits{S.}},
\bauthor{\bsnm{{Zerr}}, \binits{A.}},
\bauthor{\bsnm{{Negri}}, \binits{B.}},
\bauthor{\bsnm{{Castronuovo}}, \binits{M.}},
\bauthor{\bsnm{{Gabrielli}}, \binits{A.}},
\bauthor{\bsnm{{Bertacin}}, \binits{R.}},
\bauthor{\bsnm{{Carpentiero}}, \binits{R.}},
\bauthor{\bsnm{{Natalucci}}, \binits{S.}},
\bauthor{\bsnm{{Marliani}}, \binits{F.}},
\bauthor{\bsnm{{Cesa}}, \binits{M.}},
\bauthor{\bsnm{{Laget}}, \binits{P.}},
\bauthor{\bsnm{{Morea}}, \binits{D.}},
\bauthor{\bsnm{{Pieraccini}}, \binits{S.}},
\bauthor{\bsnm{{Radaelli}}, \binits{P.}},
\bauthor{\bsnm{{Sandri}}, \binits{P.}},
\bauthor{\bsnm{{Sarra}}, \binits{P.}},
\bauthor{\bsnm{{Cesare}}, \binits{S.}},
\bauthor{\bsnm{{Del Forno}}, \binits{F.}},
\bauthor{\bsnm{{Massa}}, \binits{E.}},
\bauthor{\bsnm{{Montabone}}, \binits{M.}},
\bauthor{\bsnm{{Mottini}}, \binits{S.}},
\bauthor{\bsnm{{Quattropani}}, \binits{D.}},
\bauthor{\bsnm{{Schillaci}}, \binits{T.}},
\bauthor{\bsnm{{Boccardo}}, \binits{R.}},
\bauthor{\bsnm{{Brando}}, \binits{R.}},
\bauthor{\bsnm{{Pandi}}, \binits{A.}},
\bauthor{\bsnm{{Baietto}}, \binits{C.}},
\bauthor{\bsnm{{Bertone}}, \binits{R.}},
\bauthor{\bsnm{{Alvarez-Herrero}}, \binits{A.}},
\bauthor{\bsnm{{Garc{\'\i}a Parejo}}, \binits{P.}},
\bauthor{\bsnm{{Cebollero}}, \binits{M.}},
\bauthor{\bsnm{{Amoruso}}, \binits{M.}},
\bauthor{\bsnm{{Centonze}}, \binits{V.}}:
\byear{2020},
\batitle{{Metis: the Solar Orbiter visible light and ultraviolet coronal
  imager}}.
\bjtitle{\aap}
\bvolume{642},
\bfpage{A10}.
\doiurl{https://doi.org/10.1051/0004-6361/201935338}.
\adsurl{2020A&A...642A..10A}.
\end{barticle}
\endbibitem

\bibitem[\protect\citeauthoryear{{Archontis} and {Hood}}{2013}]{Arc2013}
\begin{barticle}
\bauthor{\bsnm{{Archontis}}, \binits{V.}},
\bauthor{\bsnm{{Hood}}, \binits{A.W.}}:
\byear{2013},
\batitle{{A Numerical Model of Standard to Blowout Jets}}.
\bjtitle{\apjl}
\bvolume{769},
\bfpage{L21}.
\doiurl{https://doi.org/10.1088/2041-8205/769/2/L21}.
\adsurl{2013ApJ...769L..21A}.
\end{barticle}
\endbibitem

\bibitem[\protect\citeauthoryear{{Aulanier} and {Demoulin}}{1998}]{Aul1998}
\begin{barticle}
\bauthor{\bsnm{{Aulanier}}, \binits{G.}},
\bauthor{\bsnm{{Demoulin}}, \binits{P.}}:
\byear{1998},
\batitle{{3-D magnetic configurations supporting prominences. I. The natural
  presence of lateral feet}}.
\bjtitle{\aap}
\bvolume{329},
\bfpage{1125}.
\adsurl{1998A&A...329.1125A}.
\end{barticle}
\endbibitem

\bibitem[\protect\citeauthoryear{{Awasthi}, {Liu}, and {Wang}}{2019}]{Awa2019}
\begin{barticle}
\bauthor{\bsnm{{Awasthi}}, \binits{A.K.}},
\bauthor{\bsnm{{Liu}}, \binits{R.}},
\bauthor{\bsnm{{Wang}}, \binits{Y.}}:
\byear{2019},
\batitle{{Double-decker Filament Configuration Revealed by Mass Motions}}.
\bjtitle{\apj}
\bvolume{872},
\bfpage{109}.
\doiurl{https://doi.org/10.3847/1538-4357/aafdad}.
\adsurl{2019ApJ...872..109A}.
\end{barticle}
\endbibitem

\bibitem[\protect\citeauthoryear{{Berger} et~al.}{2010}]{Ber2010}
\begin{barticle}
\bauthor{\bsnm{{Berger}}, \binits{T.E.}},
\bauthor{\bsnm{{Slater}}, \binits{G.}},
\bauthor{\bsnm{{Hurlburt}}, \binits{N.}},
\bauthor{\bsnm{{Shine}}, \binits{R.}},
\bauthor{\bsnm{{Tarbell}}, \binits{T.}},
\bauthor{\bsnm{{Title}}, \binits{A.}},
\bauthor{\bsnm{{Lites}}, \binits{B.W.}},
\bauthor{\bsnm{{Okamoto}}, \binits{T.J.}},
\bauthor{\bsnm{{Ichimoto}}, \binits{K.}},
\bauthor{\bsnm{{Katsukawa}}, \binits{Y.}},
\bauthor{\bsnm{{Magara}}, \binits{T.}},
\bauthor{\bsnm{{Suematsu}}, \binits{Y.}},
\bauthor{\bsnm{{Shimizu}}, \binits{T.}}:
\byear{2010},
\batitle{{Quiescent Prominence Dynamics Observed with the Hinode Solar Optical
  Telescope. I. Turbulent Upflow Plumes}}.
\bjtitle{\apj}
\bvolume{716},
\bfpage{1288}.
\doiurl{https://doi.org/10.1088/0004-637X/716/2/1288}.
\adsurl{2010ApJ...716.1288B}.
\end{barticle}
\endbibitem

\bibitem[\protect\citeauthoryear{{Bi} et~al.}{2015}]{Bi2015}
\begin{barticle}
\bauthor{\bsnm{{Bi}}, \binits{Y.}},
\bauthor{\bsnm{{Jiang}}, \binits{Y.}},
\bauthor{\bsnm{{Yang}}, \binits{J.}},
\bauthor{\bsnm{{Xiang}}, \binits{Y.}},
\bauthor{\bsnm{{Cai}}, \binits{Y.}},
\bauthor{\bsnm{{Liu}}, \binits{W.}}:
\byear{2015},
\batitle{{Partial Eruption of a Filament with Twisting Non-uniform Fields}}.
\bjtitle{\apj}
\bvolume{805},
\bfpage{48}.
\doiurl{https://doi.org/10.1088/0004-637X/805/1/48}.
\adsurl{2015ApJ...805...48B}.
\end{barticle}
\endbibitem

\bibitem[\protect\citeauthoryear{{Bougeret} et~al.}{2008}]{Bou2008}
\begin{barticle}
\bauthor{\bsnm{{Bougeret}}, \binits{J.L.}},
\bauthor{\bsnm{{Goetz}}, \binits{K.}},
\bauthor{\bsnm{{Kaiser}}, \binits{M.L.}},
\bauthor{\bsnm{{Bale}}, \binits{S.D.}},
\bauthor{\bsnm{{Kellogg}}, \binits{P.J.}},
\bauthor{\bsnm{{Maksimovic}}, \binits{M.}},
\bauthor{\bsnm{{Monge}}, \binits{N.}},
\bauthor{\bsnm{{Monson}}, \binits{S.J.}},
\bauthor{\bsnm{{Astier}}, \binits{P.L.}},
\bauthor{\bsnm{{Davy}}, \binits{S.}},
\bauthor{\bsnm{{Dekkali}}, \binits{M.}},
\bauthor{\bsnm{{Hinze}}, \binits{J.J.}},
\bauthor{\bsnm{{Manning}}, \binits{R.E.}},
\bauthor{\bsnm{{Aguilar-Rodriguez}}, \binits{E.}},
\bauthor{\bsnm{{Bonnin}}, \binits{X.}},
\bauthor{\bsnm{{Briand}}, \binits{C.}},
\bauthor{\bsnm{{Cairns}}, \binits{I.H.}},
\bauthor{\bsnm{{Cattell}}, \binits{C.A.}},
\bauthor{\bsnm{{Cecconi}}, \binits{B.}},
\bauthor{\bsnm{{Eastwood}}, \binits{J.}},
\bauthor{\bsnm{{Ergun}}, \binits{R.E.}},
\bauthor{\bsnm{{Fainberg}}, \binits{J.}},
\bauthor{\bsnm{{Hoang}}, \binits{S.}},
\bauthor{\bsnm{{Huttunen}}, \binits{K.E.J.}},
\bauthor{\bsnm{{Krucker}}, \binits{S.}},
\bauthor{\bsnm{{Lecacheux}}, \binits{A.}},
\bauthor{\bsnm{{MacDowall}}, \binits{R.J.}},
\bauthor{\bsnm{{Macher}}, \binits{W.}},
\bauthor{\bsnm{{Mangeney}}, \binits{A.}},
\bauthor{\bsnm{{Meetre}}, \binits{C.A.}},
\bauthor{\bsnm{{Moussas}}, \binits{X.}},
\bauthor{\bsnm{{Nguyen}}, \binits{Q.N.}},
\bauthor{\bsnm{{Oswald}}, \binits{T.H.}},
\bauthor{\bsnm{{Pulupa}}, \binits{M.}},
\bauthor{\bsnm{{Reiner}}, \binits{M.J.}},
\bauthor{\bsnm{{Robinson}}, \binits{P.A.}},
\bauthor{\bsnm{{Rucker}}, \binits{H.}},
\bauthor{\bsnm{{Salem}}, \binits{C.}},
\bauthor{\bsnm{{Santolik}}, \binits{O.}},
\bauthor{\bsnm{{Silvis}}, \binits{J.M.}},
\bauthor{\bsnm{{Ullrich}}, \binits{R.}},
\bauthor{\bsnm{{Zarka}}, \binits{P.}},
\bauthor{\bsnm{{Zouganelis}}, \binits{I.}}:
\byear{2008},
\batitle{{S/WAVES: The Radio and Plasma Wave Investigation on the STEREO
  Mission}}.
\bjtitle{\ssr}
\bvolume{136},
\bfpage{487}.
\doiurl{https://doi.org/10.1007/s11214-007-9298-8}.
\adsurl{2008SSRv..136..487B}.
\end{barticle}
\endbibitem

\bibitem[\protect\citeauthoryear{{Brown}}{1971}]{Bro1971}
\begin{barticle}
\bauthor{\bsnm{{Brown}}, \binits{J.C.}}:
\byear{1971},
\batitle{{The Deduction of Energy Spectra of Non-Thermal Electrons in Flares
  from the Observed Dynamic Spectra of Hard X-Ray Bursts}}.
\bjtitle{\solphys}
\bvolume{18},
\bfpage{489}.
\doiurl{https://doi.org/10.1007/BF00149070}.
\adsurl{1971SoPh...18..489B}.
\end{barticle}
\endbibitem

\bibitem[\protect\citeauthoryear{{Brueckner} et~al.}{1995}]{Bru1995}
\begin{barticle}
\bauthor{\bsnm{{Brueckner}}, \binits{G.E.}},
\bauthor{\bsnm{{Howard}}, \binits{R.A.}},
\bauthor{\bsnm{{Koomen}}, \binits{M.J.}},
\bauthor{\bsnm{{Korendyke}}, \binits{C.M.}},
\bauthor{\bsnm{{Michels}}, \binits{D.J.}},
\bauthor{\bsnm{{Moses}}, \binits{J.D.}},
\bauthor{\bsnm{{Socker}}, \binits{D.G.}},
\bauthor{\bsnm{{Dere}}, \binits{K.P.}},
\bauthor{\bsnm{{Lamy}}, \binits{P.L.}},
\bauthor{\bsnm{{Llebaria}}, \binits{A.}},
\bauthor{\bsnm{{Bout}}, \binits{M.V.}},
\bauthor{\bsnm{{Schwenn}}, \binits{R.}},
\bauthor{\bsnm{{Simnett}}, \binits{G.M.}},
\bauthor{\bsnm{{Bedford}}, \binits{D.K.}},
\bauthor{\bsnm{{Eyles}}, \binits{C.J.}}:
\byear{1995},
\batitle{{The Large Angle Spectroscopic Coronagraph (LASCO)}}.
\bjtitle{\solphys}
\bvolume{162},
\bfpage{357}.
\doiurl{https://doi.org/10.1007/BF00733434}.
\adsurl{1995SoPh..162..357B}.
\end{barticle}
\endbibitem

\bibitem[\protect\citeauthoryear{{Byrne} et~al.}{2014}]{Byr2014}
\begin{barticle}
\bauthor{\bsnm{{Byrne}}, \binits{J.P.}},
\bauthor{\bsnm{{Morgan}}, \binits{H.}},
\bauthor{\bsnm{{Seaton}}, \binits{D.B.}},
\bauthor{\bsnm{{Bain}}, \binits{H.M.}},
\bauthor{\bsnm{{Habbal}}, \binits{S.R.}}:
\byear{2014},
\batitle{{Bridging EUV and White-Light Observations to Inspect the Initiation
  Phase of a ``Two-Stage'' Solar Eruptive Event}}.
\bjtitle{\solphys}
\bvolume{289},
\bfpage{4545}.
\doiurl{https://doi.org/10.1007/s11207-014-0585-8}.
\adsurl{2014SoPh..289.4545B}.
\end{barticle}
\endbibitem

\bibitem[\protect\citeauthoryear{{Chandra} et~al.}{2017a}]{Chan2017}
\begin{barticle}
\bauthor{\bsnm{{Chandra}}, \binits{R.}},
\bauthor{\bsnm{{Mandrini}}, \binits{C.H.}},
\bauthor{\bsnm{{Schmieder}}, \binits{B.}},
\bauthor{\bsnm{{Joshi}}, \binits{B.}},
\bauthor{\bsnm{{Cristiani}}, \binits{G.D.}},
\bauthor{\bsnm{{Cremades}}, \binits{H.}},
\bauthor{\bsnm{{Pariat}}, \binits{E.}},
\bauthor{\bsnm{{Nuevo}}, \binits{F.A.}},
\bauthor{\bsnm{{Srivastava}}, \binits{A.K.}},
\bauthor{\bsnm{{Uddin}}, \binits{W.}}:
\byear{2017}a,
\batitle{{Blowout jets and impulsive eruptive flares in a bald-patch
  topology}}.
\bjtitle{\aap}
\bvolume{598},
\bfpage{A41}.
\doiurl{https://doi.org/10.1051/0004-6361/201628984}.
\adsurl{2017A&A...598A..41C}.
\end{barticle}
\endbibitem

\bibitem[\protect\citeauthoryear{{Chandra} et~al.}{2017b}]{Cha2017}
\begin{barticle}
\bauthor{\bsnm{{Chandra}}, \binits{R.}},
\bauthor{\bsnm{{Filippov}}, \binits{B.}},
\bauthor{\bsnm{{Joshi}}, \binits{R.}},
\bauthor{\bsnm{{Schmieder}}, \binits{B.}}:
\byear{2017}b,
\batitle{{Two-Step Filament Eruption During 14 - 15 March 2015}}.
\bjtitle{\solphys}
\bvolume{292},
\bfpage{81}.
\doiurl{https://doi.org/10.1007/s11207-017-1104-5}.
\adsurl{2017SoPh..292...81C}.
\end{barticle}
\endbibitem

\bibitem[\protect\citeauthoryear{{Chen} et~al.}{2018}]{Chen2018}
\begin{barticle}
\bauthor{\bsnm{{Chen}}, \binits{H.}},
\bauthor{\bsnm{{Duan}}, \binits{Y.}},
\bauthor{\bsnm{{Yang}}, \binits{J.}},
\bauthor{\bsnm{{Yang}}, \binits{B.}},
\bauthor{\bsnm{{Dai}}, \binits{J.}}:
\byear{2018},
\batitle{{Witnessing Tether-cutting Reconnection at the Onset of a Partial
  Eruption}}.
\bjtitle{\apj}
\bvolume{869},
\bfpage{78}.
\doiurl{https://doi.org/10.3847/1538-4357/aaead1}.
\adsurl{2018ApJ...869...78C}.
\end{barticle}
\endbibitem

\bibitem[\protect\citeauthoryear{{Chen} et~al.}{2021}]{Chen2021}
\begin{barticle}
\bauthor{\bsnm{{Chen}}, \binits{J.}},
\bauthor{\bsnm{{Su}}, \binits{Y.}},
\bauthor{\bsnm{{Liu}}, \binits{R.}},
\bauthor{\bsnm{{Kliem}}, \binits{B.}},
\bauthor{\bsnm{{Zhang}}, \binits{Q.}},
\bauthor{\bsnm{{Ji}}, \binits{H.}},
\bauthor{\bsnm{{Liu}}, \binits{T.}}:
\byear{2021},
\batitle{{Partial Eruption, Confinement, and Twist Buildup and Release of a
  Double-decker Filament}}.
\bjtitle{\apj}
\bvolume{923},
\bfpage{142}.
\doiurl{https://doi.org/10.3847/1538-4357/ac2ba1}.
\adsurl{2021ApJ...923..142C}.
\end{barticle}
\endbibitem

\bibitem[\protect\citeauthoryear{{Chen} and {Shibata}}{2000}]{Chen2000}
\begin{barticle}
\bauthor{\bsnm{{Chen}}, \binits{P.F.}},
\bauthor{\bsnm{{Shibata}}, \binits{K.}}:
\byear{2000},
\batitle{{An Emerging Flux Trigger Mechanism for Coronal Mass Ejections}}.
\bjtitle{\apj}
\bvolume{545},
\bfpage{524}.
\doiurl{https://doi.org/10.1086/317803}.
\adsurl{2000ApJ...545..524C}.
\end{barticle}
\endbibitem

\bibitem[\protect\citeauthoryear{{Cheng}, {Kliem}, and
  {Ding}}{2018}]{Cheng2018}
\begin{barticle}
\bauthor{\bsnm{{Cheng}}, \binits{X.}},
\bauthor{\bsnm{{Kliem}}, \binits{B.}},
\bauthor{\bsnm{{Ding}}, \binits{M.D.}}:
\byear{2018},
\batitle{{Unambiguous Evidence of Filament Splitting-induced Partial
  Eruptions}}.
\bjtitle{\apj}
\bvolume{856},
\bfpage{48}.
\doiurl{https://doi.org/10.3847/1538-4357/aab08d}.
\adsurl{2018ApJ...856...48C}.
\end{barticle}
\endbibitem

\bibitem[\protect\citeauthoryear{{Cheng} et~al.}{2014}]{Cheng2014}
\begin{barticle}
\bauthor{\bsnm{{Cheng}}, \binits{X.}},
\bauthor{\bsnm{{Ding}}, \binits{M.D.}},
\bauthor{\bsnm{{Zhang}}, \binits{J.}},
\bauthor{\bsnm{{Sun}}, \binits{X.D.}},
\bauthor{\bsnm{{Guo}}, \binits{Y.}},
\bauthor{\bsnm{{Wang}}, \binits{Y.M.}},
\bauthor{\bsnm{{Kliem}}, \binits{B.}},
\bauthor{\bsnm{{Deng}}, \binits{Y.Y.}}:
\byear{2014},
\batitle{{Formation of a Double-decker Magnetic Flux Rope in the Sigmoidal
  Solar Active Region 11520}}.
\bjtitle{\apj}
\bvolume{789},
\bfpage{93}.
\doiurl{https://doi.org/10.1088/0004-637X/789/2/93}.
\adsurl{2014ApJ...789...93C}.
\end{barticle}
\endbibitem

\bibitem[\protect\citeauthoryear{{Dai} et~al.}{2021}]{Dai2021}
\begin{barticle}
\bauthor{\bsnm{{Dai}}, \binits{J.}},
\bauthor{\bsnm{{Zhang}}, \binits{Q.}},
\bauthor{\bsnm{{Zhang}}, \binits{Y.}},
\bauthor{\bsnm{{Xu}}, \binits{Z.}},
\bauthor{\bsnm{{Su}}, \binits{Y.}},
\bauthor{\bsnm{{Ji}}, \binits{H.}}:
\byear{2021},
\batitle{{Oscillations and Mass Draining that Lead to a Sympathetic Eruption of
  a Quiescent Filament}}.
\bjtitle{\apj}
\bvolume{923},
\bfpage{74}.
\doiurl{https://doi.org/10.3847/1538-4357/ac2d97}.
\adsurl{2021ApJ...923...74D}.
\end{barticle}
\endbibitem

\bibitem[\protect\citeauthoryear{{Dai} et~al.}{2022}]{Dai2022}
\begin{barticle}
\bauthor{\bsnm{{Dai}}, \binits{J.}},
\bauthor{\bsnm{{Li}}, \binits{Z.}},
\bauthor{\bsnm{{Wang}}, \binits{Y.}},
\bauthor{\bsnm{{Xu}}, \binits{Z.}},
\bauthor{\bsnm{{Zhang}}, \binits{Y.}},
\bauthor{\bsnm{{Li}}, \binits{L.}},
\bauthor{\bsnm{{Zhang}}, \binits{Q.}},
\bauthor{\bsnm{{Su}}, \binits{Y.}},
\bauthor{\bsnm{{Ji}}, \binits{H.}}:
\byear{2022},
\batitle{{A Partial Filament Eruption in Three Steps Induced by External
  Magnetic Reconnection}}.
\bjtitle{\apj}
\bvolume{929},
\bfpage{85}.
\doiurl{https://doi.org/10.3847/1538-4357/ac4fbe}.
\adsurl{2022ApJ...929...85D}.
\end{barticle}
\endbibitem

\bibitem[\protect\citeauthoryear{{Duan} et~al.}{2022}]{Duan2022}
\begin{barticle}
\bauthor{\bsnm{{Duan}}, \binits{Y.}},
\bauthor{\bsnm{{Shen}}, \binits{Y.}},
\bauthor{\bsnm{{Zhou}}, \binits{X.}},
\bauthor{\bsnm{{Tang}}, \binits{Z.}},
\bauthor{\bsnm{{Zhou}}, \binits{C.}},
\bauthor{\bsnm{{Tan}}, \binits{S.}}:
\byear{2022},
\batitle{{Homologous Accelerated Electron Beams, a Quasiperiodic
  Fast-propagating Wave, and a Coronal Mass Ejection Observed in One Fan-spine
  Jet}}.
\bjtitle{\apjl}
\bvolume{926},
\bfpage{L39}.
\doiurl{https://doi.org/10.3847/2041-8213/ac4df2}.
\adsurl{2022ApJ...926L..39D}.
\end{barticle}
\endbibitem

\bibitem[\protect\citeauthoryear{{Engvold}}{1998}]{Eng1998}
\begin{bchapter}
\bauthor{\bsnm{{Engvold}}, \binits{O.}}:
\byear{1998},
\bctitle{{Observations of Filament Structure and Dynamics (Review)}}.
In: \beditor{\bsnm{{Webb}}, \binits{D.F.}},
\beditor{\bsnm{{Schmieder}}, \binits{B.}},
\beditor{\bsnm{{Rust}}, \binits{D.M.}} (eds.)
\bbtitle{IAU Colloq. 167: New Perspectives on Solar Prominences},
\bsertitle{Astronomical Society of the Pacific Conference Series}
\bseriesno{150},
\bfpage{23}.
\adsurl{1998ASPC..150...23E}.
\end{bchapter}
\endbibitem

\bibitem[\protect\citeauthoryear{{Fletcher} et~al.}{2013}]{Fle2013}
\begin{barticle}
\bauthor{\bsnm{{Fletcher}}, \binits{L.}},
\bauthor{\bsnm{{Hannah}}, \binits{I.G.}},
\bauthor{\bsnm{{Hudson}}, \binits{H.S.}},
\bauthor{\bsnm{{Innes}}, \binits{D.E.}}:
\byear{2013},
\batitle{{Flare Ribbon Energetics in the Early Phase of an SDO Flare}}.
\bjtitle{\apj}
\bvolume{771},
\bfpage{104}.
\doiurl{https://doi.org/10.1088/0004-637X/771/2/104}.
\adsurl{2013ApJ...771..104F}.
\end{barticle}
\endbibitem

\bibitem[\protect\citeauthoryear{{Fox} et~al.}{2016}]{Fox2016}
\begin{barticle}
\bauthor{\bsnm{{Fox}}, \binits{N.J.}},
\bauthor{\bsnm{{Velli}}, \binits{M.C.}},
\bauthor{\bsnm{{Bale}}, \binits{S.D.}},
\bauthor{\bsnm{{Decker}}, \binits{R.}},
\bauthor{\bsnm{{Driesman}}, \binits{A.}},
\bauthor{\bsnm{{Howard}}, \binits{R.A.}},
\bauthor{\bsnm{{Kasper}}, \binits{J.C.}},
\bauthor{\bsnm{{Kinnison}}, \binits{J.}},
\bauthor{\bsnm{{Kusterer}}, \binits{M.}},
\bauthor{\bsnm{{Lario}}, \binits{D.}},
\bauthor{\bsnm{{Lockwood}}, \binits{M.K.}},
\bauthor{\bsnm{{McComas}}, \binits{D.J.}},
\bauthor{\bsnm{{Raouafi}}, \binits{N.E.}},
\bauthor{\bsnm{{Szabo}}, \binits{A.}}:
\byear{2016},
\batitle{{The Solar Probe Plus Mission: Humanity's First Visit to Our Star}}.
\bjtitle{\ssr}
\bvolume{204},
\bfpage{7}.
\doiurl{https://doi.org/10.1007/s11214-015-0211-6}.
\adsurl{2016SSRv..204....7F}.
\end{barticle}
\endbibitem

\bibitem[\protect\citeauthoryear{{Gibson}}{2018}]{Gib2018}
\begin{barticle}
\bauthor{\bsnm{{Gibson}}, \binits{S.E.}}:
\byear{2018},
\batitle{{Solar prominences: theory and models. Fleshing out the magnetic
  skeleton}}.
\bjtitle{Living Reviews in Solar Physics}
\bvolume{15},
\bfpage{7}.
\doiurl{https://doi.org/10.1007/s41116-018-0016-2}.
\adsurl{2018LRSP...15....7G}.
\end{barticle}
\endbibitem

\bibitem[\protect\citeauthoryear{{Gibson} and {Fan}}{2006}]{Gib2006}
\begin{barticle}
\bauthor{\bsnm{{Gibson}}, \binits{S.E.}},
\bauthor{\bsnm{{Fan}}, \binits{Y.}}:
\byear{2006},
\batitle{{The Partial Expulsion of a Magnetic Flux Rope}}.
\bjtitle{\apjl}
\bvolume{637},
\bfpage{L65}.
\doiurl{https://doi.org/10.1086/500452}.
\adsurl{2006ApJ...637L..65G}.
\end{barticle}
\endbibitem

\bibitem[\protect\citeauthoryear{{Gibson} and {Fan}}{2008}]{Gib2008}
\begin{barticle}
\bauthor{\bsnm{{Gibson}}, \binits{S.E.}},
\bauthor{\bsnm{{Fan}}, \binits{Y.}}:
\byear{2008},
\batitle{{Partially ejected flux ropes: Implications for interplanetary coronal
  mass ejections}}.
\bjtitle{Journal of Geophysical Research (Space Physics)}
\bvolume{113},
\bfpage{A09103}.
\doiurl{https://doi.org/10.1029/2008JA013151}.
\adsurl{2008JGRA..113.9103G}.
\end{barticle}
\endbibitem

\bibitem[\protect\citeauthoryear{{Gilbert}, {Alexander}, and
  {Liu}}{2007}]{Gil2007}
\begin{barticle}
\bauthor{\bsnm{{Gilbert}}, \binits{H.R.}},
\bauthor{\bsnm{{Alexander}}, \binits{D.}},
\bauthor{\bsnm{{Liu}}, \binits{R.}}:
\byear{2007},
\batitle{{Filament Kinking and Its Implications for Eruption and
  Re-formation}}.
\bjtitle{\solphys}
\bvolume{245},
\bfpage{287}.
\doiurl{https://doi.org/10.1007/s11207-007-9045-z}.
\adsurl{2007SoPh..245..287G}.
\end{barticle}
\endbibitem

\bibitem[\protect\citeauthoryear{{Gilbert}, {Holzer}, and
  {Burkepile}}{2001}]{Gil2001}
\begin{barticle}
\bauthor{\bsnm{{Gilbert}}, \binits{H.R.}},
\bauthor{\bsnm{{Holzer}}, \binits{T.E.}},
\bauthor{\bsnm{{Burkepile}}, \binits{J.T.}}:
\byear{2001},
\batitle{{Observational Interpretation of an Active Prominence on 1999 May 1}}.
\bjtitle{\apj}
\bvolume{549},
\bfpage{1221}.
\doiurl{https://doi.org/10.1086/319444}.
\adsurl{2001ApJ...549.1221G}.
\end{barticle}
\endbibitem

\bibitem[\protect\citeauthoryear{{Gilbert} et~al.}{2000}]{Gil2000}
\begin{barticle}
\bauthor{\bsnm{{Gilbert}}, \binits{H.R.}},
\bauthor{\bsnm{{Holzer}}, \binits{T.E.}},
\bauthor{\bsnm{{Burkepile}}, \binits{J.T.}},
\bauthor{\bsnm{{Hundhausen}}, \binits{A.J.}}:
\byear{2000},
\batitle{{Active and Eruptive Prominences and Their Relationship to Coronal
  Mass Ejections}}.
\bjtitle{\apj}
\bvolume{537},
\bfpage{503}.
\doiurl{https://doi.org/10.1086/309030}.
\adsurl{2000ApJ...537..503G}.
\end{barticle}
\endbibitem

\bibitem[\protect\citeauthoryear{{Gosain} et~al.}{2016}]{Gos2016}
\begin{barticle}
\bauthor{\bsnm{{Gosain}}, \binits{S.}},
\bauthor{\bsnm{{Filippov}}, \binits{B.}},
\bauthor{\bsnm{{Ajor Maurya}}, \binits{R.}},
\bauthor{\bsnm{{Chandra}}, \binits{R.}}:
\byear{2016},
\batitle{{Interrupted Eruption of Large Quiescent Filament Associated with a
  Halo CME}}.
\bjtitle{\apj}
\bvolume{821},
\bfpage{85}.
\doiurl{https://doi.org/10.3847/0004-637X/821/2/85}.
\adsurl{2016ApJ...821...85G}.
\end{barticle}
\endbibitem

\bibitem[\protect\citeauthoryear{{Hillier} et~al.}{2011}]{Hill2011}
\begin{barticle}
\bauthor{\bsnm{{Hillier}}, \binits{A.}},
\bauthor{\bsnm{{Isobe}}, \binits{H.}},
\bauthor{\bsnm{{Shibata}}, \binits{K.}},
\bauthor{\bsnm{{Berger}}, \binits{T.}}:
\byear{2011},
\batitle{{Numerical Simulations of the Magnetic Rayleigh-Taylor Instability in
  the Kippenhahn-Schl{\"u}ter Prominence Model}}.
\bjtitle{\apjl}
\bvolume{736},
\bfpage{L1}.
\doiurl{https://doi.org/10.1088/2041-8205/736/1/L1}.
\adsurl{2011ApJ...736L...1H}.
\end{barticle}
\endbibitem

\bibitem[\protect\citeauthoryear{{Hong} et~al.}{2011}]{Hon2011}
\begin{barticle}
\bauthor{\bsnm{{Hong}}, \binits{J.}},
\bauthor{\bsnm{{Jiang}}, \binits{Y.}},
\bauthor{\bsnm{{Zheng}}, \binits{R.}},
\bauthor{\bsnm{{Yang}}, \binits{J.}},
\bauthor{\bsnm{{Bi}}, \binits{Y.}},
\bauthor{\bsnm{{Yang}}, \binits{B.}}:
\byear{2011},
\batitle{{A Micro Coronal Mass Ejection Associated Blowout Extreme-ultraviolet
  Jet}}.
\bjtitle{\apjl}
\bvolume{738},
\bfpage{L20}.
\doiurl{https://doi.org/10.1088/2041-8205/738/2/L20}.
\adsurl{2011ApJ...738L..20H}.
\end{barticle}
\endbibitem

\bibitem[\protect\citeauthoryear{{Hong} et~al.}{2016}]{Hong2016}
\begin{barticle}
\bauthor{\bsnm{{Hong}}, \binits{J.}},
\bauthor{\bsnm{{Jiang}}, \binits{Y.}},
\bauthor{\bsnm{{Yang}}, \binits{J.}},
\bauthor{\bsnm{{Yang}}, \binits{B.}},
\bauthor{\bsnm{{Xu}}, \binits{Z.}},
\bauthor{\bsnm{{Xiang}}, \binits{Y.}}:
\byear{2016},
\batitle{{Mini-filament Eruption as the Initiation of a Jet along Coronal
  Loops}}.
\bjtitle{\apj}
\bvolume{830},
\bfpage{60}.
\doiurl{https://doi.org/10.3847/0004-637X/830/2/60}.
\adsurl{2016ApJ...830...60H}.
\end{barticle}
\endbibitem

\bibitem[\protect\citeauthoryear{{Howard} et~al.}{2008}]{How2008}
\begin{barticle}
\bauthor{\bsnm{{Howard}}, \binits{R.A.}},
\bauthor{\bsnm{{Moses}}, \binits{J.D.}},
\bauthor{\bsnm{{Vourlidas}}, \binits{A.}},
\bauthor{\bsnm{{Newmark}}, \binits{J.S.}},
\bauthor{\bsnm{{Socker}}, \binits{D.G.}},
\bauthor{\bsnm{{Plunkett}}, \binits{S.P.}},
\bauthor{\bsnm{{Korendyke}}, \binits{C.M.}},
\bauthor{\bsnm{{Cook}}, \binits{J.W.}},
\bauthor{\bsnm{{Hurley}}, \binits{A.}},
\bauthor{\bsnm{{Davila}}, \binits{J.M.}},
\bauthor{\bsnm{{Thompson}}, \binits{W.T.}},
\bauthor{\bsnm{{St Cyr}}, \binits{O.C.}},
\bauthor{\bsnm{{Mentzell}}, \binits{E.}},
\bauthor{\bsnm{{Mehalick}}, \binits{K.}},
\bauthor{\bsnm{{Lemen}}, \binits{J.R.}},
\bauthor{\bsnm{{Wuelser}}, \binits{J.P.}},
\bauthor{\bsnm{{Duncan}}, \binits{D.W.}},
\bauthor{\bsnm{{Tarbell}}, \binits{T.D.}},
\bauthor{\bsnm{{Wolfson}}, \binits{C.J.}},
\bauthor{\bsnm{{Moore}}, \binits{A.}},
\bauthor{\bsnm{{Harrison}}, \binits{R.A.}},
\bauthor{\bsnm{{Waltham}}, \binits{N.R.}},
\bauthor{\bsnm{{Lang}}, \binits{J.}},
\bauthor{\bsnm{{Davis}}, \binits{C.J.}},
\bauthor{\bsnm{{Eyles}}, \binits{C.J.}},
\bauthor{\bsnm{{Mapson-Menard}}, \binits{H.}},
\bauthor{\bsnm{{Simnett}}, \binits{G.M.}},
\bauthor{\bsnm{{Halain}}, \binits{J.P.}},
\bauthor{\bsnm{{Defise}}, \binits{J.M.}},
\bauthor{\bsnm{{Mazy}}, \binits{E.}},
\bauthor{\bsnm{{Rochus}}, \binits{P.}},
\bauthor{\bsnm{{Mercier}}, \binits{R.}},
\bauthor{\bsnm{{Ravet}}, \binits{M.F.}},
\bauthor{\bsnm{{Delmotte}}, \binits{F.}},
\bauthor{\bsnm{{Auchere}}, \binits{F.}},
\bauthor{\bsnm{{Delaboudiniere}}, \binits{J.P.}},
\bauthor{\bsnm{{Bothmer}}, \binits{V.}},
\bauthor{\bsnm{{Deutsch}}, \binits{W.}},
\bauthor{\bsnm{{Wang}}, \binits{D.}},
\bauthor{\bsnm{{Rich}}, \binits{N.}},
\bauthor{\bsnm{{Cooper}}, \binits{S.}},
\bauthor{\bsnm{{Stephens}}, \binits{V.}},
\bauthor{\bsnm{{Maahs}}, \binits{G.}},
\bauthor{\bsnm{{Baugh}}, \binits{R.}},
\bauthor{\bsnm{{McMullin}}, \binits{D.}},
\bauthor{\bsnm{{Carter}}, \binits{T.}}:
\byear{2008},
\batitle{{Sun Earth Connection Coronal and Heliospheric Investigation
  (SECCHI)}}.
\bjtitle{\ssr}
\bvolume{136},
\bfpage{67}.
\doiurl{https://doi.org/10.1007/s11214-008-9341-4}.
\adsurl{2008SSRv..136...67H}.
\end{barticle}
\endbibitem

\bibitem[\protect\citeauthoryear{{Huang} et~al.}{2018}]{Hua2018}
\begin{barticle}
\bauthor{\bsnm{{Huang}}, \binits{Z.}},
\bauthor{\bsnm{{Mou}}, \binits{C.}},
\bauthor{\bsnm{{Fu}}, \binits{H.}},
\bauthor{\bsnm{{Deng}}, \binits{L.}},
\bauthor{\bsnm{{Li}}, \binits{B.}},
\bauthor{\bsnm{{Xia}}, \binits{L.}}:
\byear{2018},
\batitle{{A Magnetic Reconnection Event in the Solar Atmosphere Driven by
  Relaxation of a Twisted Arch Filament System}}.
\bjtitle{\apjl}
\bvolume{853},
\bfpage{L26}.
\doiurl{https://doi.org/10.3847/2041-8213/aaa88c}.
\adsurl{2018ApJ...853L..26H}.
\end{barticle}
\endbibitem

\bibitem[\protect\citeauthoryear{{Ji} et~al.}{2003}]{Ji2003}
\begin{barticle}
\bauthor{\bsnm{{Ji}}, \binits{H.}},
\bauthor{\bsnm{{Wang}}, \binits{H.}},
\bauthor{\bsnm{{Schmahl}}, \binits{E.J.}},
\bauthor{\bsnm{{Moon}}, \binits{Y.-J.}},
\bauthor{\bsnm{{Jiang}}, \binits{Y.}}:
\byear{2003},
\batitle{{Observations of the Failed Eruption of a Filament}}.
\bjtitle{\apjl}
\bvolume{595},
\bfpage{L135}.
\doiurl{https://doi.org/10.1086/378178}.
\adsurl{2003ApJ...595L.135J}.
\end{barticle}
\endbibitem

\bibitem[\protect\citeauthoryear{{Joshi} et~al.}{2014}]{Jos2014}
\begin{barticle}
\bauthor{\bsnm{{Joshi}}, \binits{N.C.}},
\bauthor{\bsnm{{Srivastava}}, \binits{A.K.}},
\bauthor{\bsnm{{Filippov}}, \binits{B.}},
\bauthor{\bsnm{{Kayshap}}, \binits{P.}},
\bauthor{\bsnm{{Uddin}}, \binits{W.}},
\bauthor{\bsnm{{Chandra}}, \binits{R.}},
\bauthor{\bsnm{{Prasad Choudhary}}, \binits{D.}},
\bauthor{\bsnm{{Dwivedi}}, \binits{B.N.}}:
\byear{2014},
\batitle{{Confined Partial Filament Eruption and its Reformation within a
  Stable Magnetic Flux Rope}}.
\bjtitle{\apj}
\bvolume{787},
\bfpage{11}.
\doiurl{https://doi.org/10.1088/0004-637X/787/1/11}.
\adsurl{2014ApJ...787...11J}.
\end{barticle}
\endbibitem

\bibitem[\protect\citeauthoryear{{Joshi} et~al.}{2018}]{Jos2018}
\begin{barticle}
\bauthor{\bsnm{{Joshi}}, \binits{N.C.}},
\bauthor{\bsnm{{Nishizuka}}, \binits{N.}},
\bauthor{\bsnm{{Filippov}}, \binits{B.}},
\bauthor{\bsnm{{Magara}}, \binits{T.}},
\bauthor{\bsnm{{Tlatov}}, \binits{A.G.}}:
\byear{2018},
\batitle{{Flux rope breaking and formation of a rotating blowout jet}}.
\bjtitle{\mnras}
\bvolume{476},
\bfpage{1286}.
\doiurl{https://doi.org/10.1093/mnras/sty322}.
\adsurl{2018MNRAS.476.1286J}.
\end{barticle}
\endbibitem

\bibitem[\protect\citeauthoryear{{Joshi} et~al.}{2022}]{Jos2022}
\begin{barticle}
\bauthor{\bsnm{{Joshi}}, \binits{R.}},
\bauthor{\bsnm{{Mandrini}}, \binits{C.H.}},
\bauthor{\bsnm{{Chandra}}, \binits{R.}},
\bauthor{\bsnm{{Schmieder}}, \binits{B.}},
\bauthor{\bsnm{{Cristiani}}, \binits{G.D.}},
\bauthor{\bsnm{{Mac Cormack}}, \binits{C.}},
\bauthor{\bsnm{{D{\'e}moulin}}, \binits{P.}},
\bauthor{\bsnm{{Cremades}}, \binits{H.}}:
\byear{2022},
\batitle{{Analysis of the Evolution of a Multi-Ribbon Flare and Failed Filament
  Eruption}}.
\bjtitle{\solphys}
\bvolume{297},
\bfpage{81}.
\doiurl{https://doi.org/10.1007/s11207-022-02021-5}.
\adsurl{2022SoPh..297...81J}.
\end{barticle}
\endbibitem

\bibitem[\protect\citeauthoryear{{Kaiser} et~al.}{2008}]{Kai2008}
\begin{barticle}
\bauthor{\bsnm{{Kaiser}}, \binits{M.L.}},
\bauthor{\bsnm{{Kucera}}, \binits{T.A.}},
\bauthor{\bsnm{{Davila}}, \binits{J.M.}},
\bauthor{\bsnm{{St. Cyr}}, \binits{O.C.}},
\bauthor{\bsnm{{Guhathakurta}}, \binits{M.}},
\bauthor{\bsnm{{Christian}}, \binits{E.}}:
\byear{2008},
\batitle{{The STEREO Mission: An Introduction}}.
\bjtitle{\ssr}
\bvolume{136},
\bfpage{5}.
\doiurl{https://doi.org/10.1007/s11214-007-9277-0}.
\adsurl{2008SSRv..136....5K}.
\end{barticle}
\endbibitem

\bibitem[\protect\citeauthoryear{{Karpen} et~al.}{2005}]{Kar2005}
\begin{barticle}
\bauthor{\bsnm{{Karpen}}, \binits{J.T.}},
\bauthor{\bsnm{{Tanner}}, \binits{S.E.M.}},
\bauthor{\bsnm{{Antiochos}}, \binits{S.K.}},
\bauthor{\bsnm{{DeVore}}, \binits{C.R.}}:
\byear{2005},
\batitle{{Prominence Formation by Thermal Nonequilibrium in the Sheared-Arcade
  Model}}.
\bjtitle{\apj}
\bvolume{635},
\bfpage{1319}.
\doiurl{https://doi.org/10.1086/497531}.
\adsurl{2005ApJ...635.1319K}.
\end{barticle}
\endbibitem

\bibitem[\protect\citeauthoryear{{Kliem} et~al.}{2014}]{Kli2014}
\begin{barticle}
\bauthor{\bsnm{{Kliem}}, \binits{B.}},
\bauthor{\bsnm{{T{\"o}r{\"o}k}}, \binits{T.}},
\bauthor{\bsnm{{Titov}}, \binits{V.S.}},
\bauthor{\bsnm{{Lionello}}, \binits{R.}},
\bauthor{\bsnm{{Linker}}, \binits{J.A.}},
\bauthor{\bsnm{{Liu}}, \binits{R.}},
\bauthor{\bsnm{{Liu}}, \binits{C.}},
\bauthor{\bsnm{{Wang}}, \binits{H.}}:
\byear{2014},
\batitle{{Slow Rise and Partial Eruption of a Double-decker Filament. II. A
  Double Flux Rope Model}}.
\bjtitle{\apj}
\bvolume{792},
\bfpage{107}.
\doiurl{https://doi.org/10.1088/0004-637X/792/2/107}.
\adsurl{2014ApJ...792..107K}.
\end{barticle}
\endbibitem

\bibitem[\protect\citeauthoryear{{Krucker} et~al.}{2011}]{Kru2011}
\begin{barticle}
\bauthor{\bsnm{{Krucker}}, \binits{S.}},
\bauthor{\bsnm{{Kontar}}, \binits{E.P.}},
\bauthor{\bsnm{{Christe}}, \binits{S.}},
\bauthor{\bsnm{{Glesener}}, \binits{L.}},
\bauthor{\bsnm{{Lin}}, \binits{R.P.}}:
\byear{2011},
\batitle{{Electron Acceleration Associated with Solar Jets}}.
\bjtitle{\apj}
\bvolume{742},
\bfpage{82}.
\doiurl{https://doi.org/10.1088/0004-637X/742/2/82}.
\adsurl{2011ApJ...742...82K}.
\end{barticle}
\endbibitem

\bibitem[\protect\citeauthoryear{{Labrosse} et~al.}{2010}]{Lab2010}
\begin{barticle}
\bauthor{\bsnm{{Labrosse}}, \binits{N.}},
\bauthor{\bsnm{{Heinzel}}, \binits{P.}},
\bauthor{\bsnm{{Vial}}, \binits{J.-C.}},
\bauthor{\bsnm{{Kucera}}, \binits{T.}},
\bauthor{\bsnm{{Parenti}}, \binits{S.}},
\bauthor{\bsnm{{Gun{\'a}r}}, \binits{S.}},
\bauthor{\bsnm{{Schmieder}}, \binits{B.}},
\bauthor{\bsnm{{Kilper}}, \binits{G.}}:
\byear{2010},
\batitle{{Physics of Solar Prominences: I{\textemdash}Spectral Diagnostics and
  Non-LTE Modelling}}.
\bjtitle{\ssr}
\bvolume{151},
\bfpage{243}.
\doiurl{https://doi.org/10.1007/s11214-010-9630-6}.
\adsurl{2010SSRv..151..243L}.
\end{barticle}
\endbibitem

\bibitem[\protect\citeauthoryear{{Lemen} et~al.}{2012}]{Lem2012}
\begin{barticle}
\bauthor{\bsnm{{Lemen}}, \binits{J.R.}},
\bauthor{\bsnm{{Title}}, \binits{A.M.}},
\bauthor{\bsnm{{Akin}}, \binits{D.J.}},
\bauthor{\bsnm{{Boerner}}, \binits{P.F.}},
\bauthor{\bsnm{{Chou}}, \binits{C.}},
\bauthor{\bsnm{{Drake}}, \binits{J.F.}},
\bauthor{\bsnm{{Duncan}}, \binits{D.W.}},
\bauthor{\bsnm{{Edwards}}, \binits{C.G.}},
\bauthor{\bsnm{{Friedlaender}}, \binits{F.M.}},
\bauthor{\bsnm{{Heyman}}, \binits{G.F.}},
\bauthor{\bsnm{{Hurlburt}}, \binits{N.E.}},
\bauthor{\bsnm{{Katz}}, \binits{N.L.}},
\bauthor{\bsnm{{Kushner}}, \binits{G.D.}},
\bauthor{\bsnm{{Levay}}, \binits{M.}},
\bauthor{\bsnm{{Lindgren}}, \binits{R.W.}},
\bauthor{\bsnm{{Mathur}}, \binits{D.P.}},
\bauthor{\bsnm{{McFeaters}}, \binits{E.L.}},
\bauthor{\bsnm{{Mitchell}}, \binits{S.}},
\bauthor{\bsnm{{Rehse}}, \binits{R.A.}},
\bauthor{\bsnm{{Schrijver}}, \binits{C.J.}},
\bauthor{\bsnm{{Springer}}, \binits{L.A.}},
\bauthor{\bsnm{{Stern}}, \binits{R.A.}},
\bauthor{\bsnm{{Tarbell}}, \binits{T.D.}},
\bauthor{\bsnm{{Wuelser}}, \binits{J.-P.}},
\bauthor{\bsnm{{Wolfson}}, \binits{C.J.}},
\bauthor{\bsnm{{Yanari}}, \binits{C.}},
\bauthor{\bsnm{{Bookbinder}}, \binits{J.A.}},
\bauthor{\bsnm{{Cheimets}}, \binits{P.N.}},
\bauthor{\bsnm{{Caldwell}}, \binits{D.}},
\bauthor{\bsnm{{Deluca}}, \binits{E.E.}},
\bauthor{\bsnm{{Gates}}, \binits{R.}},
\bauthor{\bsnm{{Golub}}, \binits{L.}},
\bauthor{\bsnm{{Park}}, \binits{S.}},
\bauthor{\bsnm{{Podgorski}}, \binits{W.A.}},
\bauthor{\bsnm{{Bush}}, \binits{R.I.}},
\bauthor{\bsnm{{Scherrer}}, \binits{P.H.}},
\bauthor{\bsnm{{Gummin}}, \binits{M.A.}},
\bauthor{\bsnm{{Smith}}, \binits{P.}},
\bauthor{\bsnm{{Auker}}, \binits{G.}},
\bauthor{\bsnm{{Jerram}}, \binits{P.}},
\bauthor{\bsnm{{Pool}}, \binits{P.}},
\bauthor{\bsnm{{Soufli}}, \binits{R.}},
\bauthor{\bsnm{{Windt}}, \binits{D.L.}},
\bauthor{\bsnm{{Beardsley}}, \binits{S.}},
\bauthor{\bsnm{{Clapp}}, \binits{M.}},
\bauthor{\bsnm{{Lang}}, \binits{J.}},
\bauthor{\bsnm{{Waltham}}, \binits{N.}}:
\byear{2012},
\batitle{{The Atmospheric Imaging Assembly (AIA) on the Solar Dynamics
  Observatory (SDO)}}.
\bjtitle{\solphys}
\bvolume{275},
\bfpage{17}.
\doiurl{https://doi.org/10.1007/s11207-011-9776-8}.
\adsurl{2012SoPh..275...17L}.
\end{barticle}
\endbibitem

\bibitem[\protect\citeauthoryear{{Li} et~al.}{2016}]{Li2016}
\begin{barticle}
\bauthor{\bsnm{{Li}}, \binits{L.}},
\bauthor{\bsnm{{Zhang}}, \binits{J.}},
\bauthor{\bsnm{{Peter}}, \binits{H.}},
\bauthor{\bsnm{{Priest}}, \binits{E.}},
\bauthor{\bsnm{{Chen}}, \binits{H.}},
\bauthor{\bsnm{{Guo}}, \binits{L.}},
\bauthor{\bsnm{{Chen}}, \binits{F.}},
\bauthor{\bsnm{{Mackay}}, \binits{D.}}:
\byear{2016},
\batitle{{Magnetic reconnection between a solar filament and nearby coronal
  loops}}.
\bjtitle{Nature Physics}
\bvolume{12},
\bfpage{847}.
\doiurl{https://doi.org/10.1038/nphys3768}.
\adsurl{2016NatPh..12..847L}.
\end{barticle}
\endbibitem

\bibitem[\protect\citeauthoryear{{Li} et~al.}{2018}]{Li2018}
\begin{barticle}
\bauthor{\bsnm{{Li}}, \binits{T.}},
\bauthor{\bsnm{{Yang}}, \binits{S.}},
\bauthor{\bsnm{{Zhang}}, \binits{Q.}},
\bauthor{\bsnm{{Hou}}, \binits{Y.}},
\bauthor{\bsnm{{Zhang}}, \binits{J.}}:
\byear{2018},
\batitle{{Two Episodes of Magnetic Reconnections during a Confined
  Circular-ribbon Flare}}.
\bjtitle{\apj}
\bvolume{859},
\bfpage{122}.
\doiurl{https://doi.org/10.3847/1538-4357/aabe84}.
\adsurl{2018ApJ...859..122L}.
\end{barticle}
\endbibitem

\bibitem[\protect\citeauthoryear{{Li} et~al.}{2015}]{Li2015}
\begin{barticle}
\bauthor{\bsnm{{Li}}, \binits{X.}},
\bauthor{\bsnm{{Yang}}, \binits{S.}},
\bauthor{\bsnm{{Chen}}, \binits{H.}},
\bauthor{\bsnm{{Li}}, \binits{T.}},
\bauthor{\bsnm{{Zhang}}, \binits{J.}}:
\byear{2015},
\batitle{{Trigger of a Blowout Jet in a Solar Coronal Mass Ejection Associated
  with a Flare}}.
\bjtitle{\apjl}
\bvolume{814},
\bfpage{L13}.
\doiurl{https://doi.org/10.1088/2041-8205/814/1/L13}.
\adsurl{2015ApJ...814L..13L}.
\end{barticle}
\endbibitem

\bibitem[\protect\citeauthoryear{{Lin}, {Martin}, and
  {Engvold}}{2008}]{Lin2008}
\begin{bchapter}
\bauthor{\bsnm{{Lin}}, \binits{Y.}},
\bauthor{\bsnm{{Martin}}, \binits{S.F.}},
\bauthor{\bsnm{{Engvold}}, \binits{O.}}:
\byear{2008},
\bctitle{{Filament Substructures and their Interrelation}}.
In: \beditor{\bsnm{{Howe}}, \binits{R.}},
\beditor{\bsnm{{Komm}}, \binits{R.W.}},
\beditor{\bsnm{{Balasubramaniam}}, \binits{K.S.}},
\beditor{\bsnm{{Petrie}}, \binits{G.J.D.}} (eds.)
\bbtitle{Subsurface and Atmospheric Influences on Solar Activity},
\bsertitle{Astronomical Society of the Pacific Conference Series}
\bseriesno{383},
\bfpage{235}.
\adsurl{2008ASPC..383..235L}.
\end{bchapter}
\endbibitem

\bibitem[\protect\citeauthoryear{{Lin} et~al.}{2005}]{Lin2005}
\begin{barticle}
\bauthor{\bsnm{{Lin}}, \binits{Y.}},
\bauthor{\bsnm{{Engvold}}, \binits{O.}},
\bauthor{\bsnm{{Rouppe van der Voort}}, \binits{L.}},
\bauthor{\bsnm{{Wiik}}, \binits{J.E.}},
\bauthor{\bsnm{{Berger}}, \binits{T.E.}}:
\byear{2005},
\batitle{{Thin Threads of Solar Filaments}}.
\bjtitle{\solphys}
\bvolume{226},
\bfpage{239}.
\doiurl{https://doi.org/10.1007/s11207-005-6876-3}.
\adsurl{2005SoPh..226..239L}.
\end{barticle}
\endbibitem

\bibitem[\protect\citeauthoryear{{Liu}, {Alexander}, and
  {Gilbert}}{2007}]{Liu2007}
\begin{barticle}
\bauthor{\bsnm{{Liu}}, \binits{R.}},
\bauthor{\bsnm{{Alexander}}, \binits{D.}},
\bauthor{\bsnm{{Gilbert}}, \binits{H.R.}}:
\byear{2007},
\batitle{{Kink-induced Catastrophe in a Coronal Eruption}}.
\bjtitle{\apj}
\bvolume{661},
\bfpage{1260}.
\doiurl{https://doi.org/10.1086/513269}.
\adsurl{2007ApJ...661.1260L}.
\end{barticle}
\endbibitem

\bibitem[\protect\citeauthoryear{{Liu}, {Chen}, and {Wang}}{2018}]{Liu2018}
\begin{barticle}
\bauthor{\bsnm{{Liu}}, \binits{R.}},
\bauthor{\bsnm{{Chen}}, \binits{J.}},
\bauthor{\bsnm{{Wang}}, \binits{Y.}}:
\byear{2018},
\batitle{{Disintegration of an eruptive filament via interactions with
  quasi-separatrix layers}}.
\bjtitle{Science China Physics, Mechanics, and Astronomy}
\bvolume{61},
\bfpage{69611}.
\doiurl{https://doi.org/10.1007/s11433-017-9147-x}.
\adsurl{2018SCPMA..61f9611L}.
\end{barticle}
\endbibitem

\bibitem[\protect\citeauthoryear{{Liu} et~al.}{2012}]{Liu2012a}
\begin{barticle}
\bauthor{\bsnm{{Liu}}, \binits{R.}},
\bauthor{\bsnm{{Kliem}}, \binits{B.}},
\bauthor{\bsnm{{T{\"o}r{\"o}k}}, \binits{T.}},
\bauthor{\bsnm{{Liu}}, \binits{C.}},
\bauthor{\bsnm{{Titov}}, \binits{V.S.}},
\bauthor{\bsnm{{Lionello}}, \binits{R.}},
\bauthor{\bsnm{{Linker}}, \binits{J.A.}},
\bauthor{\bsnm{{Wang}}, \binits{H.}}:
\byear{2012},
\batitle{{Slow Rise and Partial Eruption of a Double-decker Filament. I.
  Observations and Interpretation}}.
\bjtitle{\apj}
\bvolume{756},
\bfpage{59}.
\doiurl{https://doi.org/10.1088/0004-637X/756/1/59}.
\adsurl{2012ApJ...756...59L}.
\end{barticle}
\endbibitem

\bibitem[\protect\citeauthoryear{{Liu}, {Berger}, and {Low}}{2012}]{Liu2012b}
\begin{barticle}
\bauthor{\bsnm{{Liu}}, \binits{W.}},
\bauthor{\bsnm{{Berger}}, \binits{T.E.}},
\bauthor{\bsnm{{Low}}, \binits{B.C.}}:
\byear{2012},
\batitle{{First SDO/AIA Observation of Solar Prominence Formation Following an
  Eruption: Magnetic Dips and Sustained Condensation and Drainage}}.
\bjtitle{\apjl}
\bvolume{745},
\bfpage{L21}.
\doiurl{https://doi.org/10.1088/2041-8205/745/2/L21}.
\adsurl{2012ApJ...745L..21L}.
\end{barticle}
\endbibitem

\bibitem[\protect\citeauthoryear{{Liu} et~al.}{2006}]{Liu2006}
\begin{barticle}
\bauthor{\bsnm{{Liu}}, \binits{W.}},
\bauthor{\bsnm{{Liu}}, \binits{S.}},
\bauthor{\bsnm{{Jiang}}, \binits{Y.W.}},
\bauthor{\bsnm{{Petrosian}}, \binits{V.}}:
\byear{2006},
\batitle{{RHESSI Observation of Chromospheric Evaporation}}.
\bjtitle{\apj}
\bvolume{649},
\bfpage{1124}.
\doiurl{https://doi.org/10.1086/506268}.
\adsurl{2006ApJ...649.1124L}.
\end{barticle}
\endbibitem

\bibitem[\protect\citeauthoryear{{Liu} et~al.}{2009}]{Liu2009}
\begin{barticle}
\bauthor{\bsnm{{Liu}}, \binits{Y.}},
\bauthor{\bsnm{{Su}}, \binits{J.}},
\bauthor{\bsnm{{Xu}}, \binits{Z.}},
\bauthor{\bsnm{{Lin}}, \binits{H.}},
\bauthor{\bsnm{{Shibata}}, \binits{K.}},
\bauthor{\bsnm{{Kurokawa}}, \binits{H.}}:
\byear{2009},
\batitle{{New Observation of Failed Filament Eruptions: The Influence of
  Asymmetric Coronal Background Fields on Solar Eruptions}}.
\bjtitle{\apjl}
\bvolume{696},
\bfpage{L70}.
\doiurl{https://doi.org/10.1088/0004-637X/696/1/L70}.
\adsurl{2009ApJ...696L..70L}.
\end{barticle}
\endbibitem

\bibitem[\protect\citeauthoryear{{Luna} et~al.}{2018}]{Luna2018}
\begin{barticle}
\bauthor{\bsnm{{Luna}}, \binits{M.}},
\bauthor{\bsnm{{Karpen}}, \binits{J.}},
\bauthor{\bsnm{{Ballester}}, \binits{J.L.}},
\bauthor{\bsnm{{Muglach}}, \binits{K.}},
\bauthor{\bsnm{{Terradas}}, \binits{J.}},
\bauthor{\bsnm{{Kucera}}, \binits{T.}},
\bauthor{\bsnm{{Gilbert}}, \binits{H.}}:
\byear{2018},
\batitle{{GONG Catalog of Solar Filament Oscillations Near Solar Maximum}}.
\bjtitle{\apjs}
\bvolume{236},
\bfpage{35}.
\doiurl{https://doi.org/10.3847/1538-4365/aabde7}.
\adsurl{2018ApJS..236...35L}.
\end{barticle}
\endbibitem

\bibitem[\protect\citeauthoryear{{Mackay} et~al.}{2010}]{Mac2010}
\begin{barticle}
\bauthor{\bsnm{{Mackay}}, \binits{D.H.}},
\bauthor{\bsnm{{Karpen}}, \binits{J.T.}},
\bauthor{\bsnm{{Ballester}}, \binits{J.L.}},
\bauthor{\bsnm{{Schmieder}}, \binits{B.}},
\bauthor{\bsnm{{Aulanier}}, \binits{G.}}:
\byear{2010},
\batitle{{Physics of Solar Prominences: II{\textemdash}Magnetic Structure and
  Dynamics}}.
\bjtitle{\ssr}
\bvolume{151},
\bfpage{333}.
\doiurl{https://doi.org/10.1007/s11214-010-9628-0}.
\adsurl{2010SSRv..151..333M}.
\end{barticle}
\endbibitem

\bibitem[\protect\citeauthoryear{{Martin}}{1998}]{Mar1998}
\begin{barticle}
\bauthor{\bsnm{{Martin}}, \binits{S.F.}}:
\byear{1998},
\batitle{{Conditions for the Formation and Maintenance of Filaments (Invited
  Review)}}.
\bjtitle{\solphys}
\bvolume{182},
\bfpage{107}.
\doiurl{https://doi.org/10.1023/A:1005026814076}.
\adsurl{1998SoPh..182..107M}.
\end{barticle}
\endbibitem

\bibitem[\protect\citeauthoryear{{Masson}, {Antiochos}, and
  {DeVore}}{2013}]{Mas2013}
\begin{barticle}
\bauthor{\bsnm{{Masson}}, \binits{S.}},
\bauthor{\bsnm{{Antiochos}}, \binits{S.K.}},
\bauthor{\bsnm{{DeVore}}, \binits{C.R.}}:
\byear{2013},
\batitle{{A Model for the Escape of Solar-flare-accelerated Particles}}.
\bjtitle{\apj}
\bvolume{771},
\bfpage{82}.
\doiurl{https://doi.org/10.1088/0004-637X/771/2/82}.
\adsurl{2013ApJ...771...82M}.
\end{barticle}
\endbibitem

\bibitem[\protect\citeauthoryear{{McCauley} et~al.}{2015}]{Mc2015}
\begin{barticle}
\bauthor{\bsnm{{McCauley}}, \binits{P.I.}},
\bauthor{\bsnm{{Su}}, \binits{Y.N.}},
\bauthor{\bsnm{{Schanche}}, \binits{N.}},
\bauthor{\bsnm{{Evans}}, \binits{K.E.}},
\bauthor{\bsnm{{Su}}, \binits{C.}},
\bauthor{\bsnm{{McKillop}}, \binits{S.}},
\bauthor{\bsnm{{Reeves}}, \binits{K.K.}}:
\byear{2015},
\batitle{{Prominence and Filament Eruptions Observed by the Solar Dynamics
  Observatory: Statistical Properties, Kinematics, and Online Catalog}}.
\bjtitle{\solphys}
\bvolume{290},
\bfpage{1703}.
\doiurl{https://doi.org/10.1007/s11207-015-0699-7}.
\adsurl{2015SoPh..290.1703M}.
\end{barticle}
\endbibitem

\bibitem[\protect\citeauthoryear{{Meegan} et~al.}{2009}]{Mee2009}
\begin{barticle}
\bauthor{\bsnm{{Meegan}}, \binits{C.}},
\bauthor{\bsnm{{Lichti}}, \binits{G.}},
\bauthor{\bsnm{{Bhat}}, \binits{P.N.}},
\bauthor{\bsnm{{Bissaldi}}, \binits{E.}},
\bauthor{\bsnm{{Briggs}}, \binits{M.S.}},
\bauthor{\bsnm{{Connaughton}}, \binits{V.}},
\bauthor{\bsnm{{Diehl}}, \binits{R.}},
\bauthor{\bsnm{{Fishman}}, \binits{G.}},
\bauthor{\bsnm{{Greiner}}, \binits{J.}},
\bauthor{\bsnm{{Hoover}}, \binits{A.S.}},
\bauthor{\bsnm{{van der Horst}}, \binits{A.J.}},
\bauthor{\bsnm{{von Kienlin}}, \binits{A.}},
\bauthor{\bsnm{{Kippen}}, \binits{R.M.}},
\bauthor{\bsnm{{Kouveliotou}}, \binits{C.}},
\bauthor{\bsnm{{McBreen}}, \binits{S.}},
\bauthor{\bsnm{{Paciesas}}, \binits{W.S.}},
\bauthor{\bsnm{{Preece}}, \binits{R.}},
\bauthor{\bsnm{{Steinle}}, \binits{H.}},
\bauthor{\bsnm{{Wallace}}, \binits{M.S.}},
\bauthor{\bsnm{{Wilson}}, \binits{R.B.}},
\bauthor{\bsnm{{Wilson-Hodge}}, \binits{C.}}:
\byear{2009},
\batitle{{The Fermi Gamma-ray Burst Monitor}}.
\bjtitle{\apj}
\bvolume{702},
\bfpage{791}.
\doiurl{https://doi.org/10.1088/0004-637X/702/1/791}.
\adsurl{2009ApJ...702..791M}.
\end{barticle}
\endbibitem

\bibitem[\protect\citeauthoryear{{Monga} et~al.}{2021}]{Mon2021}
\begin{barticle}
\bauthor{\bsnm{{Monga}}, \binits{A.}},
\bauthor{\bsnm{{Sharma}}, \binits{R.}},
\bauthor{\bsnm{{Liu}}, \binits{J.}},
\bauthor{\bsnm{{Cid}}, \binits{C.}},
\bauthor{\bsnm{{Uddin}}, \binits{W.}},
\bauthor{\bsnm{{Chandra}}, \binits{R.}},
\bauthor{\bsnm{{Erd{\'e}lyi}}, \binits{R.}}:
\byear{2021},
\batitle{{On the partial eruption of a bifurcated solar filament structure}}.
\bjtitle{\mnras}
\bvolume{500},
\bfpage{684}.
\doiurl{https://doi.org/10.1093/mnras/staa2902}.
\adsurl{2021MNRAS.500..684M}.
\end{barticle}
\endbibitem

\bibitem[\protect\citeauthoryear{{Moreno-Insertis}, {Galsgaard}, and
  {Ugarte-Urra}}{2008}]{Mor2008}
\begin{barticle}
\bauthor{\bsnm{{Moreno-Insertis}}, \binits{F.}},
\bauthor{\bsnm{{Galsgaard}}, \binits{K.}},
\bauthor{\bsnm{{Ugarte-Urra}}, \binits{I.}}:
\byear{2008},
\batitle{{Jets in Coronal Holes: Hinode Observations and Three-dimensional
  Computer Modeling}}.
\bjtitle{\apjl}
\bvolume{673},
\bfpage{L211}.
\doiurl{https://doi.org/10.1086/527560}.
\adsurl{2008ApJ...673L.211M}.
\end{barticle}
\endbibitem

\bibitem[\protect\citeauthoryear{{M{\"u}ller} et~al.}{2020}]{Mul2020}
\begin{barticle}
\bauthor{\bsnm{{M{\"u}ller}}, \binits{D.}},
\bauthor{\bsnm{{St. Cyr}}, \binits{O.C.}},
\bauthor{\bsnm{{Zouganelis}}, \binits{I.}},
\bauthor{\bsnm{{Gilbert}}, \binits{H.R.}},
\bauthor{\bsnm{{Marsden}}, \binits{R.}},
\bauthor{\bsnm{{Nieves-Chinchilla}}, \binits{T.}},
\bauthor{\bsnm{{Antonucci}}, \binits{E.}},
\bauthor{\bsnm{{Auch{\`e}re}}, \binits{F.}},
\bauthor{\bsnm{{Berghmans}}, \binits{D.}},
\bauthor{\bsnm{{Horbury}}, \binits{T.S.}},
\bauthor{\bsnm{{Howard}}, \binits{R.A.}},
\bauthor{\bsnm{{Krucker}}, \binits{S.}},
\bauthor{\bsnm{{Maksimovic}}, \binits{M.}},
\bauthor{\bsnm{{Owen}}, \binits{C.J.}},
\bauthor{\bsnm{{Rochus}}, \binits{P.}},
\bauthor{\bsnm{{Rodriguez-Pacheco}}, \binits{J.}},
\bauthor{\bsnm{{Romoli}}, \binits{M.}},
\bauthor{\bsnm{{Solanki}}, \binits{S.K.}},
\bauthor{\bsnm{{Bruno}}, \binits{R.}},
\bauthor{\bsnm{{Carlsson}}, \binits{M.}},
\bauthor{\bsnm{{Fludra}}, \binits{A.}},
\bauthor{\bsnm{{Harra}}, \binits{L.}},
\bauthor{\bsnm{{Hassler}}, \binits{D.M.}},
\bauthor{\bsnm{{Livi}}, \binits{S.}},
\bauthor{\bsnm{{Louarn}}, \binits{P.}},
\bauthor{\bsnm{{Peter}}, \binits{H.}},
\bauthor{\bsnm{{Sch{\"u}hle}}, \binits{U.}},
\bauthor{\bsnm{{Teriaca}}, \binits{L.}},
\bauthor{\bsnm{{del Toro Iniesta}}, \binits{J.C.}},
\bauthor{\bsnm{{Wimmer-Schweingruber}}, \binits{R.F.}},
\bauthor{\bsnm{{Marsch}}, \binits{E.}},
\bauthor{\bsnm{{Velli}}, \binits{M.}},
\bauthor{\bsnm{{De Groof}}, \binits{A.}},
\bauthor{\bsnm{{Walsh}}, \binits{A.}},
\bauthor{\bsnm{{Williams}}, \binits{D.}}:
\byear{2020},
\batitle{{The Solar Orbiter mission. Science overview}}.
\bjtitle{\aap}
\bvolume{642},
\bfpage{A1}.
\doiurl{https://doi.org/10.1051/0004-6361/202038467}.
\adsurl{2020A&A...642A...1M}.
\end{barticle}
\endbibitem

\bibitem[\protect\citeauthoryear{{Ning} et~al.}{2009}]{Ning2009}
\begin{barticle}
\bauthor{\bsnm{{Ning}}, \binits{Z.}},
\bauthor{\bsnm{{Cao}}, \binits{W.}},
\bauthor{\bsnm{{Okamoto}}, \binits{T.J.}},
\bauthor{\bsnm{{Ichimoto}}, \binits{K.}},
\bauthor{\bsnm{{Qu}}, \binits{Z.Q.}}:
\byear{2009},
\batitle{{Small-scale oscillations in a quiescent prominence observed by
  HINODE/SOT. Prominence oscillations}}.
\bjtitle{\aap}
\bvolume{499},
\bfpage{595}.
\doiurl{https://doi.org/10.1051/0004-6361/200810853}.
\adsurl{2009A&A...499..595N}.
\end{barticle}
\endbibitem

\bibitem[\protect\citeauthoryear{{Nistic{\`o}} et~al.}{2009}]{Nis2009}
\begin{barticle}
\bauthor{\bsnm{{Nistic{\`o}}}, \binits{G.}},
\bauthor{\bsnm{{Bothmer}}, \binits{V.}},
\bauthor{\bsnm{{Patsourakos}}, \binits{S.}},
\bauthor{\bsnm{{Zimbardo}}, \binits{G.}}:
\byear{2009},
\batitle{{Characteristics of EUV Coronal Jets Observed with STEREO/SECCHI}}.
\bjtitle{\solphys}
\bvolume{259},
\bfpage{87}.
\doiurl{https://doi.org/10.1007/s11207-009-9424-8}.
\adsurl{2009SoPh..259...87N}.
\end{barticle}
\endbibitem

\bibitem[\protect\citeauthoryear{{Okamoto} et~al.}{2007}]{Oka2007}
\begin{barticle}
\bauthor{\bsnm{{Okamoto}}, \binits{T.J.}},
\bauthor{\bsnm{{Tsuneta}}, \binits{S.}},
\bauthor{\bsnm{{Berger}}, \binits{T.E.}},
\bauthor{\bsnm{{Ichimoto}}, \binits{K.}},
\bauthor{\bsnm{{Katsukawa}}, \binits{Y.}},
\bauthor{\bsnm{{Lites}}, \binits{B.W.}},
\bauthor{\bsnm{{Nagata}}, \binits{S.}},
\bauthor{\bsnm{{Shibata}}, \binits{K.}},
\bauthor{\bsnm{{Shimizu}}, \binits{T.}},
\bauthor{\bsnm{{Shine}}, \binits{R.A.}},
\bauthor{\bsnm{{Suematsu}}, \binits{Y.}},
\bauthor{\bsnm{{Tarbell}}, \binits{T.D.}},
\bauthor{\bsnm{{Title}}, \binits{A.M.}}:
\byear{2007},
\batitle{{Coronal Transverse Magnetohydrodynamic Waves in a Solar Prominence}}.
\bjtitle{Science}
\bvolume{318},
\bfpage{1577}.
\doiurl{https://doi.org/10.1126/science.1145447}.
\adsurl{2007Sci...318.1577O}.
\end{barticle}
\endbibitem

\bibitem[\protect\citeauthoryear{{O'Kane} et~al.}{2021}]{Oka2021}
\begin{barticle}
\bauthor{\bsnm{{O'Kane}}, \binits{J.}},
\bauthor{\bsnm{{Green}}, \binits{L.M.}},
\bauthor{\bsnm{{Davies}}, \binits{E.E.}},
\bauthor{\bsnm{{M{\"o}stl}}, \binits{C.}},
\bauthor{\bsnm{{Hinterreiter}}, \binits{J.}},
\bauthor{\bsnm{{Freiherr von Forstner}}, \binits{J.L.}},
\bauthor{\bsnm{{Weiss}}, \binits{A.J.}},
\bauthor{\bsnm{{Long}}, \binits{D.M.}},
\bauthor{\bsnm{{Amerstorfer}}, \binits{T.}}:
\byear{2021},
\batitle{{Solar origins of a strong stealth CME detected by Solar Orbiter}}.
\bjtitle{\aap}
\bvolume{656},
\bfpage{L6}.
\doiurl{https://doi.org/10.1051/0004-6361/202140622}.
\adsurl{2021A&A...656L...6O}.
\end{barticle}
\endbibitem

\bibitem[\protect\citeauthoryear{{Ouyang} et~al.}{2017}]{OY2017}
\begin{barticle}
\bauthor{\bsnm{{Ouyang}}, \binits{Y.}},
\bauthor{\bsnm{{Zhou}}, \binits{Y.H.}},
\bauthor{\bsnm{{Chen}}, \binits{P.F.}},
\bauthor{\bsnm{{Fang}}, \binits{C.}}:
\byear{2017},
\batitle{{Chirality and Magnetic Configurations of Solar Filaments}}.
\bjtitle{\apj}
\bvolume{835},
\bfpage{94}.
\doiurl{https://doi.org/10.3847/1538-4357/835/1/94}.
\adsurl{2017ApJ...835...94O}.
\end{barticle}
\endbibitem

\bibitem[\protect\citeauthoryear{{Pan} et~al.}{2021}]{Pan2021}
\begin{barticle}
\bauthor{\bsnm{{Pan}}, \binits{H.}},
\bauthor{\bsnm{{Liu}}, \binits{R.}},
\bauthor{\bsnm{{Gou}}, \binits{T.}},
\bauthor{\bsnm{{Kliem}}, \binits{B.}},
\bauthor{\bsnm{{Su}}, \binits{Y.}},
\bauthor{\bsnm{{Chen}}, \binits{J.}},
\bauthor{\bsnm{{Wang}}, \binits{Y.}}:
\byear{2021},
\batitle{{Pre-eruption Splitting of the Double-decker Structure in a Solar
  Filament}}.
\bjtitle{\apj}
\bvolume{909},
\bfpage{32}.
\doiurl{https://doi.org/10.3847/1538-4357/abda4e}.
\adsurl{2021ApJ...909...32P}.
\end{barticle}
\endbibitem

\bibitem[\protect\citeauthoryear{{Panesar}, {Sterling}, and
  {Moore}}{2016}]{Pan2016b}
\begin{barticle}
\bauthor{\bsnm{{Panesar}}, \binits{N.K.}},
\bauthor{\bsnm{{Sterling}}, \binits{A.C.}},
\bauthor{\bsnm{{Moore}}, \binits{R.L.}}:
\byear{2016},
\batitle{{Homologous Jet-driven Coronal Mass Ejections from Solar Active Region
  12192}}.
\bjtitle{\apjl}
\bvolume{822},
\bfpage{L23}.
\doiurl{https://doi.org/10.3847/2041-8205/822/2/L23}.
\adsurl{2016ApJ...822L..23P}.
\end{barticle}
\endbibitem

\bibitem[\protect\citeauthoryear{{Parenti}}{2014}]{Par2014}
\begin{barticle}
\bauthor{\bsnm{{Parenti}}, \binits{S.}}:
\byear{2014},
\batitle{{Solar Prominences: Observations}}.
\bjtitle{Living Reviews in Solar Physics}
\bvolume{11},
\bfpage{1}.
\doiurl{https://doi.org/10.12942/lrsp-2014-1}.
\adsurl{2014LRSP...11....1P}.
\end{barticle}
\endbibitem

\bibitem[\protect\citeauthoryear{{Patsourakos}, {Vourlidas}, and
  {Stenborg}}{2010}]{Pats2010}
\begin{barticle}
\bauthor{\bsnm{{Patsourakos}}, \binits{S.}},
\bauthor{\bsnm{{Vourlidas}}, \binits{A.}},
\bauthor{\bsnm{{Stenborg}}, \binits{G.}}:
\byear{2010},
\batitle{{The Genesis of an Impulsive Coronal Mass Ejection Observed at
  Ultra-high Cadence by AIA on SDO}}.
\bjtitle{\apjl}
\bvolume{724},
\bfpage{L188}.
\doiurl{https://doi.org/10.1088/2041-8205/724/2/L188}.
\adsurl{2010ApJ...724L.188P}.
\end{barticle}
\endbibitem

\bibitem[\protect\citeauthoryear{{Pesnell}, {Thompson}, and
  {Chamberlin}}{2012}]{Pes2012}
\begin{barticle}
\bauthor{\bsnm{{Pesnell}}, \binits{W.D.}},
\bauthor{\bsnm{{Thompson}}, \binits{B.J.}},
\bauthor{\bsnm{{Chamberlin}}, \binits{P.C.}}:
\byear{2012},
\batitle{{The Solar Dynamics Observatory (SDO)}}.
\bjtitle{\solphys}
\bvolume{275},
\bfpage{3}.
\doiurl{https://doi.org/10.1007/s11207-011-9841-3}.
\adsurl{2012SoPh..275....3P}.
\end{barticle}
\endbibitem

\bibitem[\protect\citeauthoryear{{Poisson} et~al.}{2020}]{Poi2020}
\begin{barticle}
\bauthor{\bsnm{{Poisson}}, \binits{M.}},
\bauthor{\bsnm{{Bustos}}, \binits{C.}},
\bauthor{\bsnm{{L{\'o}pez Fuentes}}, \binits{M.}},
\bauthor{\bsnm{{Mandrini}}, \binits{C.H.}},
\bauthor{\bsnm{{Cristiani}}, \binits{G.D.}}:
\byear{2020},
\batitle{{Two successive partial mini-filament confined ejections}}.
\bjtitle{Advances in Space Research}
\bvolume{65},
\bfpage{1629}.
\doiurl{https://doi.org/10.1016/j.asr.2019.09.026}.
\adsurl{2020AdSpR..65.1629P}.
\end{barticle}
\endbibitem

\bibitem[\protect\citeauthoryear{{Priest}, {Hood}, and {Anzer}}{1989}]{Pri1989}
\begin{barticle}
\bauthor{\bsnm{{Priest}}, \binits{E.R.}},
\bauthor{\bsnm{{Hood}}, \binits{A.W.}},
\bauthor{\bsnm{{Anzer}}, \binits{U.}}:
\byear{1989},
\batitle{{A Twisted Flux-Tube Model for Solar Prominences. I. General
  Properties}}.
\bjtitle{\apj}
\bvolume{344},
\bfpage{1010}.
\doiurl{https://doi.org/10.1086/167868}.
\adsurl{1989ApJ...344.1010P}.
\end{barticle}
\endbibitem

\bibitem[\protect\citeauthoryear{{Schmieder}, {D{\'e}moulin}, and
  {Aulanier}}{2013}]{Sch2013}
\begin{barticle}
\bauthor{\bsnm{{Schmieder}}, \binits{B.}},
\bauthor{\bsnm{{D{\'e}moulin}}, \binits{P.}},
\bauthor{\bsnm{{Aulanier}}, \binits{G.}}:
\byear{2013},
\batitle{{Solar filament eruptions and their physical role in triggering
  coronal mass ejections}}.
\bjtitle{Advances in Space Research}
\bvolume{51},
\bfpage{1967}.
\doiurl{https://doi.org/10.1016/j.asr.2012.12.026}.
\adsurl{2013AdSpR..51.1967S}.
\end{barticle}
\endbibitem

\bibitem[\protect\citeauthoryear{{Schmieder} et~al.}{2014}]{Sch2014}
\begin{barticle}
\bauthor{\bsnm{{Schmieder}}, \binits{B.}},
\bauthor{\bsnm{{Tian}}, \binits{H.}},
\bauthor{\bsnm{{Kucera}}, \binits{T.}},
\bauthor{\bsnm{{L{\'o}pez Ariste}}, \binits{A.}},
\bauthor{\bsnm{{Mein}}, \binits{N.}},
\bauthor{\bsnm{{Mein}}, \binits{P.}},
\bauthor{\bsnm{{Dalmasse}}, \binits{K.}},
\bauthor{\bsnm{{Golub}}, \binits{L.}}:
\byear{2014},
\batitle{{Open questions on prominences from coordinated observations by IRIS,
  Hinode, SDO/AIA, THEMIS, and the Meudon/MSDP}}.
\bjtitle{\aap}
\bvolume{569},
\bfpage{A85}.
\doiurl{https://doi.org/10.1051/0004-6361/201423922}.
\adsurl{2014A&A...569A..85S}.
\end{barticle}
\endbibitem

\bibitem[\protect\citeauthoryear{{Shen}, {Liu}, and {Su}}{2012}]{She2012}
\begin{barticle}
\bauthor{\bsnm{{Shen}}, \binits{Y.}},
\bauthor{\bsnm{{Liu}}, \binits{Y.}},
\bauthor{\bsnm{{Su}}, \binits{J.}}:
\byear{2012},
\batitle{{Sympathetic Partial and Full Filament Eruptions Observed in One Solar
  Breakout Event}}.
\bjtitle{\apj}
\bvolume{750},
\bfpage{12}.
\doiurl{https://doi.org/10.1088/0004-637X/750/1/12}.
\adsurl{2012ApJ...750...12S}.
\end{barticle}
\endbibitem

\bibitem[\protect\citeauthoryear{{Shen} et~al.}{2012}]{Shen2012}
\begin{barticle}
\bauthor{\bsnm{{Shen}}, \binits{Y.}},
\bauthor{\bsnm{{Liu}}, \binits{Y.}},
\bauthor{\bsnm{{Su}}, \binits{J.}},
\bauthor{\bsnm{{Deng}}, \binits{Y.}}:
\byear{2012},
\batitle{{On a Coronal Blowout Jet: The First Observation of a Simultaneously
  Produced Bubble-like CME and a Jet-like CME in a Solar Event}}.
\bjtitle{\apj}
\bvolume{745},
\bfpage{164}.
\doiurl{https://doi.org/10.1088/0004-637X/745/2/164}.
\adsurl{2012ApJ...745..164S}.
\end{barticle}
\endbibitem

\bibitem[\protect\citeauthoryear{{Shibata} et~al.}{1992}]{Shi1992}
\begin{barticle}
\bauthor{\bsnm{{Shibata}}, \binits{K.}},
\bauthor{\bsnm{{Ishido}}, \binits{Y.}},
\bauthor{\bsnm{{Acton}}, \binits{L.W.}},
\bauthor{\bsnm{{Strong}}, \binits{K.T.}},
\bauthor{\bsnm{{Hirayama}}, \binits{T.}},
\bauthor{\bsnm{{Uchida}}, \binits{Y.}},
\bauthor{\bsnm{{McAllister}}, \binits{A.H.}},
\bauthor{\bsnm{{Matsumoto}}, \binits{R.}},
\bauthor{\bsnm{{Tsuneta}}, \binits{S.}},
\bauthor{\bsnm{{Shimizu}}, \binits{T.}},
\bauthor{\bsnm{{Hara}}, \binits{H.}},
\bauthor{\bsnm{{Sakurai}}, \binits{T.}},
\bauthor{\bsnm{{Ichimoto}}, \binits{K.}},
\bauthor{\bsnm{{Nishino}}, \binits{Y.}},
\bauthor{\bsnm{{Ogawara}}, \binits{Y.}}:
\byear{1992},
\batitle{{Observations of X-Ray Jets with the YOHKOH Soft X-Ray Telescope}}.
\bjtitle{\pasj}
\bvolume{44},
\bfpage{L173}.
\adsurl{1992PASJ...44L.173S}.
\end{barticle}
\endbibitem

\bibitem[\protect\citeauthoryear{{Song}, {Li}, and {Chen}}{2022}]{Song2022}
\begin{barticle}
\bauthor{\bsnm{{Song}}, \binits{H.}},
\bauthor{\bsnm{{Li}}, \binits{L.}},
\bauthor{\bsnm{{Chen}}, \binits{Y.}}:
\byear{2022},
\batitle{{Toward a Unified Explanation for the Three-part Structure of Solar
  Coronal Mass Ejections}}.
\bjtitle{\apj}
\bvolume{933},
\bfpage{68}.
\doiurl{https://doi.org/10.3847/1538-4357/ac7239}.
\adsurl{2022ApJ...933...68S}.
\end{barticle}
\endbibitem

\bibitem[\protect\citeauthoryear{{Srivastava}, {Mishra}, and
  {Jel{\'\i}nek}}{2021}]{Sri2021}
\begin{barticle}
\bauthor{\bsnm{{Srivastava}}, \binits{A.K.}},
\bauthor{\bsnm{{Mishra}}, \binits{S.K.}},
\bauthor{\bsnm{{Jel{\'\i}nek}}, \binits{P.}}:
\byear{2021},
\batitle{{The Prominence Driven Forced Reconnection in the Solar Corona and
  Associated Plasma Dynamics}}.
\bjtitle{\apj}
\bvolume{920},
\bfpage{18}.
\doiurl{https://doi.org/10.3847/1538-4357/ac1519}.
\adsurl{2021ApJ...920...18S}.
\end{barticle}
\endbibitem

\bibitem[\protect\citeauthoryear{{Sterling} et~al.}{2015}]{Ster2015}
\begin{barticle}
\bauthor{\bsnm{{Sterling}}, \binits{A.C.}},
\bauthor{\bsnm{{Moore}}, \binits{R.L.}},
\bauthor{\bsnm{{Falconer}}, \binits{D.A.}},
\bauthor{\bsnm{{Adams}}, \binits{M.}}:
\byear{2015},
\batitle{{Small-scale filament eruptions as the driver of X-ray jets in solar
  coronal holes}}.
\bjtitle{\nat}
\bvolume{523},
\bfpage{437}.
\doiurl{https://doi.org/10.1038/nature14556}.
\adsurl{2015Natur.523..437S}.
\end{barticle}
\endbibitem

\bibitem[\protect\citeauthoryear{{Su} et~al.}{2012a}]{Su2012b}
\begin{barticle}
\bauthor{\bsnm{{Su}}, \binits{Y.}},
\bauthor{\bsnm{{Dennis}}, \binits{B.R.}},
\bauthor{\bsnm{{Holman}}, \binits{G.D.}},
\bauthor{\bsnm{{Wang}}, \binits{T.}},
\bauthor{\bsnm{{Chamberlin}}, \binits{P.C.}},
\bauthor{\bsnm{{Savage}}, \binits{S.}},
\bauthor{\bsnm{{Veronig}}, \binits{A.}}:
\byear{2012}a,
\batitle{{Observations of a Two-stage Solar Eruptive Event (SEE): Evidence for
  Secondary Heating}}.
\bjtitle{\apjl}
\bvolume{746},
\bfpage{L5}.
\doiurl{https://doi.org/10.1088/2041-8205/746/1/L5}.
\adsurl{2012ApJ...746L...5S}.
\end{barticle}
\endbibitem

\bibitem[\protect\citeauthoryear{{Su} et~al.}{2012b}]{Su2012}
\begin{barticle}
\bauthor{\bsnm{{Su}}, \binits{Y.}},
\bauthor{\bsnm{{Wang}}, \binits{T.}},
\bauthor{\bsnm{{Veronig}}, \binits{A.}},
\bauthor{\bsnm{{Temmer}}, \binits{M.}},
\bauthor{\bsnm{{Gan}}, \binits{W.}}:
\byear{2012}b,
\batitle{{Solar Magnetized ``Tornadoes:'' Relation to Filaments}}.
\bjtitle{\apjl}
\bvolume{756},
\bfpage{L41}.
\doiurl{https://doi.org/10.1088/2041-8205/756/2/L41}.
\adsurl{2012ApJ...756L..41S}.
\end{barticle}
\endbibitem

\bibitem[\protect\citeauthoryear{{Terradas} et~al.}{2015}]{Ter2015}
\begin{barticle}
\bauthor{\bsnm{{Terradas}}, \binits{J.}},
\bauthor{\bsnm{{Soler}}, \binits{R.}},
\bauthor{\bsnm{{Luna}}, \binits{M.}},
\bauthor{\bsnm{{Oliver}}, \binits{R.}},
\bauthor{\bsnm{{Ballester}}, \binits{J.L.}}:
\byear{2015},
\batitle{{Morphology and Dynamics of Solar Prominences from 3D MHD
  Simulations}}.
\bjtitle{\apj}
\bvolume{799},
\bfpage{94}.
\doiurl{https://doi.org/10.1088/0004-637X/799/1/94}.
\adsurl{2015ApJ...799...94T}.
\end{barticle}
\endbibitem

\bibitem[\protect\citeauthoryear{{Tian} et~al.}{2015}]{Tia2015}
\begin{barticle}
\bauthor{\bsnm{{Tian}}, \binits{H.}},
\bauthor{\bsnm{{Young}}, \binits{P.R.}},
\bauthor{\bsnm{{Reeves}}, \binits{K.K.}},
\bauthor{\bsnm{{Chen}}, \binits{B.}},
\bauthor{\bsnm{{Liu}}, \binits{W.}},
\bauthor{\bsnm{{McKillop}}, \binits{S.}}:
\byear{2015},
\batitle{{Temporal Evolution of Chromospheric Evaporation: Case Studies of the
  M1.1 Flare on 2014 September 6 and X1.6 Flare on 2014 September 10}}.
\bjtitle{\apj}
\bvolume{811},
\bfpage{139}.
\doiurl{https://doi.org/10.1088/0004-637X/811/2/139}.
\adsurl{2015ApJ...811..139T}.
\end{barticle}
\endbibitem

\bibitem[\protect\citeauthoryear{{T{\"o}r{\"o}k} and {Kliem}}{2005}]{Tor2005}
\begin{barticle}
\bauthor{\bsnm{{T{\"o}r{\"o}k}}, \binits{T.}},
\bauthor{\bsnm{{Kliem}}, \binits{B.}}:
\byear{2005},
\batitle{{Confined and Ejective Eruptions of Kink-unstable Flux Ropes}}.
\bjtitle{\apjl}
\bvolume{630},
\bfpage{L97}.
\doiurl{https://doi.org/10.1086/462412}.
\adsurl{2005ApJ...630L..97T}.
\end{barticle}
\endbibitem

\bibitem[\protect\citeauthoryear{{Tripathi} et~al.}{2009}]{Tri2009}
\begin{barticle}
\bauthor{\bsnm{{Tripathi}}, \binits{D.}},
\bauthor{\bsnm{{Gibson}}, \binits{S.E.}},
\bauthor{\bsnm{{Qiu}}, \binits{J.}},
\bauthor{\bsnm{{Fletcher}}, \binits{L.}},
\bauthor{\bsnm{{Liu}}, \binits{R.}},
\bauthor{\bsnm{{Gilbert}}, \binits{H.}},
\bauthor{\bsnm{{Mason}}, \binits{H.E.}}:
\byear{2009},
\batitle{{Partially-erupting prominences: a comparison between observations and
  model-predicted observables}}.
\bjtitle{\aap}
\bvolume{498},
\bfpage{295}.
\doiurl{https://doi.org/10.1051/0004-6361/200809801}.
\adsurl{2009A&A...498..295T}.
\end{barticle}
\endbibitem

\bibitem[\protect\citeauthoryear{{Vourlidas} et~al.}{2016}]{Vour2016}
\begin{barticle}
\bauthor{\bsnm{{Vourlidas}}, \binits{A.}},
\bauthor{\bsnm{{Howard}}, \binits{R.A.}},
\bauthor{\bsnm{{Plunkett}}, \binits{S.P.}},
\bauthor{\bsnm{{Korendyke}}, \binits{C.M.}},
\bauthor{\bsnm{{Thernisien}}, \binits{A.F.R.}},
\bauthor{\bsnm{{Wang}}, \binits{D.}},
\bauthor{\bsnm{{Rich}}, \binits{N.}},
\bauthor{\bsnm{{Carter}}, \binits{M.T.}},
\bauthor{\bsnm{{Chua}}, \binits{D.H.}},
\bauthor{\bsnm{{Socker}}, \binits{D.G.}},
\bauthor{\bsnm{{Linton}}, \binits{M.G.}},
\bauthor{\bsnm{{Morrill}}, \binits{J.S.}},
\bauthor{\bsnm{{Lynch}}, \binits{S.}},
\bauthor{\bsnm{{Thurn}}, \binits{A.}},
\bauthor{\bsnm{{Van Duyne}}, \binits{P.}},
\bauthor{\bsnm{{Hagood}}, \binits{R.}},
\bauthor{\bsnm{{Clifford}}, \binits{G.}},
\bauthor{\bsnm{{Grey}}, \binits{P.J.}},
\bauthor{\bsnm{{Velli}}, \binits{M.}},
\bauthor{\bsnm{{Liewer}}, \binits{P.C.}},
\bauthor{\bsnm{{Hall}}, \binits{J.R.}},
\bauthor{\bsnm{{DeJong}}, \binits{E.M.}},
\bauthor{\bsnm{{Mikic}}, \binits{Z.}},
\bauthor{\bsnm{{Rochus}}, \binits{P.}},
\bauthor{\bsnm{{Mazy}}, \binits{E.}},
\bauthor{\bsnm{{Bothmer}}, \binits{V.}},
\bauthor{\bsnm{{Rodmann}}, \binits{J.}}:
\byear{2016},
\batitle{{The Wide-Field Imager for Solar Probe Plus (WISPR)}}.
\bjtitle{\ssr}
\bvolume{204},
\bfpage{83}.
\doiurl{https://doi.org/10.1007/s11214-014-0114-y}.
\adsurl{2016SSRv..204...83V}.
\end{barticle}
\endbibitem

\bibitem[\protect\citeauthoryear{{Vourlidas} et~al.}{2017}]{Vou2017}
\begin{barticle}
\bauthor{\bsnm{{Vourlidas}}, \binits{A.}},
\bauthor{\bsnm{{Balmaceda}}, \binits{L.A.}},
\bauthor{\bsnm{{Stenborg}}, \binits{G.}},
\bauthor{\bsnm{{Dal Lago}}, \binits{A.}}:
\byear{2017},
\batitle{{Multi-viewpoint Coronal Mass Ejection Catalog Based on STEREO COR2
  Observations}}.
\bjtitle{\apj}
\bvolume{838},
\bfpage{141}.
\doiurl{https://doi.org/10.3847/1538-4357/aa67f0}.
\adsurl{2017ApJ...838..141V}.
\end{barticle}
\endbibitem

\bibitem[\protect\citeauthoryear{{Wang} et~al.}{2019}]{Wang2019}
\begin{barticle}
\bauthor{\bsnm{{Wang}}, \binits{J.}},
\bauthor{\bsnm{{Yan}}, \binits{X.}},
\bauthor{\bsnm{{Guo}}, \binits{Q.}},
\bauthor{\bsnm{{Kong}}, \binits{D.}},
\bauthor{\bsnm{{Xue}}, \binits{Z.}},
\bauthor{\bsnm{{Yang}}, \binits{L.}},
\bauthor{\bsnm{{Li}}, \binits{Q.}}:
\byear{2019},
\batitle{{Formation and material supply of an active-region filament associated
  with newly emerging flux}}.
\bjtitle{\mnras}
\bvolume{488},
\bfpage{3794}.
\doiurl{https://doi.org/10.1093/mnras/stz1935}.
\adsurl{2019MNRAS.488.3794W}.
\end{barticle}
\endbibitem

\bibitem[\protect\citeauthoryear{{Wang} et~al.}{1998}]{Wang1998}
\begin{barticle}
\bauthor{\bsnm{{Wang}}, \binits{Y.-M.}},
\bauthor{\bsnm{{Sheeley}}, \binits{J.} \bsuffix{N.~R.}},
\bauthor{\bsnm{{Socker}}, \binits{D.G.}},
\bauthor{\bsnm{{Howard}}, \binits{R.A.}},
\bauthor{\bsnm{{Brueckner}}, \binits{G.E.}},
\bauthor{\bsnm{{Michels}}, \binits{D.J.}},
\bauthor{\bsnm{{Moses}}, \binits{D.}},
\bauthor{\bsnm{{St. Cyr}}, \binits{O.C.}},
\bauthor{\bsnm{{Llebaria}}, \binits{A.}},
\bauthor{\bsnm{{Delaboudini{\`e}re}}, \binits{J.-P.}}:
\byear{1998},
\batitle{{Observations of Correlated White-Light and Extreme-Ultraviolet Jets
  from Polar Coronal Holes}}.
\bjtitle{\apj}
\bvolume{508},
\bfpage{899}.
\doiurl{https://doi.org/10.1086/306450}.
\adsurl{1998ApJ...508..899W}.
\end{barticle}
\endbibitem

\bibitem[\protect\citeauthoryear{{Wei} et~al.}{2021}]{Wei2021}
\begin{barticle}
\bauthor{\bsnm{{Wei}}, \binits{Y.}},
\bauthor{\bsnm{{Chen}}, \binits{B.}},
\bauthor{\bsnm{{Yu}}, \binits{S.}},
\bauthor{\bsnm{{Wang}}, \binits{H.}},
\bauthor{\bsnm{{Jing}}, \binits{J.}},
\bauthor{\bsnm{{Gary}}, \binits{D.E.}}:
\byear{2021},
\batitle{{Coronal Magnetic Field Measurements along a Partially Erupting
  Filament in a Solar Flare}}.
\bjtitle{\apj}
\bvolume{923},
\bfpage{213}.
\doiurl{https://doi.org/10.3847/1538-4357/ac2f99}.
\adsurl{2021ApJ...923..213W}.
\end{barticle}
\endbibitem

\bibitem[\protect\citeauthoryear{{Woods} et~al.}{2011}]{Woo2011}
\begin{barticle}
\bauthor{\bsnm{{Woods}}, \binits{T.N.}},
\bauthor{\bsnm{{Hock}}, \binits{R.}},
\bauthor{\bsnm{{Eparvier}}, \binits{F.}},
\bauthor{\bsnm{{Jones}}, \binits{A.R.}},
\bauthor{\bsnm{{Chamberlin}}, \binits{P.C.}},
\bauthor{\bsnm{{Klimchuk}}, \binits{J.A.}},
\bauthor{\bsnm{{Didkovsky}}, \binits{L.}},
\bauthor{\bsnm{{Judge}}, \binits{D.}},
\bauthor{\bsnm{{Mariska}}, \binits{J.}},
\bauthor{\bsnm{{Warren}}, \binits{H.}},
\bauthor{\bsnm{{Schrijver}}, \binits{C.J.}},
\bauthor{\bsnm{{Webb}}, \binits{D.F.}},
\bauthor{\bsnm{{Bailey}}, \binits{S.}},
\bauthor{\bsnm{{Tobiska}}, \binits{W.K.}}:
\byear{2011},
\batitle{{New Solar Extreme-ultraviolet Irradiance Observations during
  Flares}}.
\bjtitle{\apj}
\bvolume{739},
\bfpage{59}.
\doiurl{https://doi.org/10.1088/0004-637X/739/2/59}.
\adsurl{2011ApJ...739...59W}.
\end{barticle}
\endbibitem

\bibitem[\protect\citeauthoryear{{Wuelser} et~al.}{2004}]{Wue2004}
\begin{bchapter}
\bauthor{\bsnm{{Wuelser}}, \binits{J.-P.}},
\bauthor{\bsnm{{Lemen}}, \binits{J.R.}},
\bauthor{\bsnm{{Tarbell}}, \binits{T.D.}},
\bauthor{\bsnm{{Wolfson}}, \binits{C.J.}},
\bauthor{\bsnm{{Cannon}}, \binits{J.C.}},
\bauthor{\bsnm{{Carpenter}}, \binits{B.A.}},
\bauthor{\bsnm{{Duncan}}, \binits{D.W.}},
\bauthor{\bsnm{{Gradwohl}}, \binits{G.S.}},
\bauthor{\bsnm{{Meyer}}, \binits{S.B.}},
\bauthor{\bsnm{{Moore}}, \binits{A.S.}},
\bauthor{\bsnm{{Navarro}}, \binits{R.L.}},
\bauthor{\bsnm{{Pearson}}, \binits{J.D.}},
\bauthor{\bsnm{{Rossi}}, \binits{G.R.}},
\bauthor{\bsnm{{Springer}}, \binits{L.A.}},
\bauthor{\bsnm{{Howard}}, \binits{R.A.}},
\bauthor{\bsnm{{Moses}}, \binits{J.D.}},
\bauthor{\bsnm{{Newmark}}, \binits{J.S.}},
\bauthor{\bsnm{{Delaboudiniere}}, \binits{J.-P.}},
\bauthor{\bsnm{{Artzner}}, \binits{G.E.}},
\bauthor{\bsnm{{Auchere}}, \binits{F.}},
\bauthor{\bsnm{{Bougnet}}, \binits{M.}},
\bauthor{\bsnm{{Bouyries}}, \binits{P.}},
\bauthor{\bsnm{{Bridou}}, \binits{F.}},
\bauthor{\bsnm{{Clotaire}}, \binits{J.-Y.}},
\bauthor{\bsnm{{Colas}}, \binits{G.}},
\bauthor{\bsnm{{Delmotte}}, \binits{F.}},
\bauthor{\bsnm{{Jerome}}, \binits{A.}},
\bauthor{\bsnm{{Lamare}}, \binits{M.}},
\bauthor{\bsnm{{Mercier}}, \binits{R.}},
\bauthor{\bsnm{{Mullot}}, \binits{M.}},
\bauthor{\bsnm{{Ravet}}, \binits{M.-F.}},
\bauthor{\bsnm{{Song}}, \binits{X.}},
\bauthor{\bsnm{{Bothmer}}, \binits{V.}},
\bauthor{\bsnm{{Deutsch}}, \binits{W.}}:
\byear{2004},
\bctitle{{EUVI: the STEREO-SECCHI extreme ultraviolet imager}}.
In: \beditor{\bsnm{{Fineschi}}, \binits{S.}},
\beditor{\bsnm{{Gummin}}, \binits{M.A.}} (eds.)
\bbtitle{Telescopes and Instrumentation for Solar Astrophysics},
\bsertitle{Society of Photo-Optical Instrumentation Engineers (SPIE) Conference
  Series}
\bseriesno{5171},
\bfpage{111}.
\doiurl{https://doi.org/10.1117/12.506877}.
\adsurl{2004SPIE.5171..111W}.
\end{bchapter}
\endbibitem

\bibitem[\protect\citeauthoryear{{Wyper}, {DeVore}, and
  {Antiochos}}{2018}]{Wyp2018}
\begin{barticle}
\bauthor{\bsnm{{Wyper}}, \binits{P.F.}},
\bauthor{\bsnm{{DeVore}}, \binits{C.R.}},
\bauthor{\bsnm{{Antiochos}}, \binits{S.K.}}:
\byear{2018},
\batitle{{A Breakout Model for Solar Coronal Jets with Filaments}}.
\bjtitle{\apj}
\bvolume{852},
\bfpage{98}.
\doiurl{https://doi.org/10.3847/1538-4357/aa9ffc}.
\adsurl{2018ApJ...852...98W}.
\end{barticle}
\endbibitem

\bibitem[\protect\citeauthoryear{{Xue} et~al.}{2016}]{Xue2016}
\begin{barticle}
\bauthor{\bsnm{{Xue}}, \binits{Z.}},
\bauthor{\bsnm{{Yan}}, \binits{X.}},
\bauthor{\bsnm{{Cheng}}, \binits{X.}},
\bauthor{\bsnm{{Yang}}, \binits{L.}},
\bauthor{\bsnm{{Su}}, \binits{Y.}},
\bauthor{\bsnm{{Kliem}}, \binits{B.}},
\bauthor{\bsnm{{Zhang}}, \binits{J.}},
\bauthor{\bsnm{{Liu}}, \binits{Z.}},
\bauthor{\bsnm{{Bi}}, \binits{Y.}},
\bauthor{\bsnm{{Xiang}}, \binits{Y.}},
\bauthor{\bsnm{{Yang}}, \binits{K.}},
\bauthor{\bsnm{{Zhao}}, \binits{L.}}:
\byear{2016},
\batitle{{Observing the release of twist by magnetic reconnection in a solar
  filament eruption}}.
\bjtitle{Nature Communications}
\bvolume{7},
\bfpage{11837}.
\doiurl{https://doi.org/10.1038/ncomms11837}.
\adsurl{2016NatCo...711837X}.
\end{barticle}
\endbibitem

\bibitem[\protect\citeauthoryear{{Yan} et~al.}{2015}]{Yan2015}
\begin{barticle}
\bauthor{\bsnm{{Yan}}, \binits{X.L.}},
\bauthor{\bsnm{{Xue}}, \binits{Z.K.}},
\bauthor{\bsnm{{Pan}}, \binits{G.M.}},
\bauthor{\bsnm{{Wang}}, \binits{J.C.}},
\bauthor{\bsnm{{Xiang}}, \binits{Y.Y.}},
\bauthor{\bsnm{{Kong}}, \binits{D.F.}},
\bauthor{\bsnm{{Yang}}, \binits{L.H.}}:
\byear{2015},
\batitle{{The Formation and Magnetic Structures of Active-region Filaments
  Observed by NVST, SDO, and Hinode}}.
\bjtitle{\apjs}
\bvolume{219},
\bfpage{17}.
\doiurl{https://doi.org/10.1088/0067-0049/219/2/17}.
\adsurl{2015ApJS..219...17Y}.
\end{barticle}
\endbibitem

\bibitem[\protect\citeauthoryear{{Zhang} and {Ji}}{2014}]{Zha2014}
\begin{barticle}
\bauthor{\bsnm{{Zhang}}, \binits{Q.M.}},
\bauthor{\bsnm{{Ji}}, \binits{H.S.}}:
\byear{2014},
\batitle{{A swirling flare-related EUV jet}}.
\bjtitle{\aap}
\bvolume{561},
\bfpage{A134}.
\doiurl{https://doi.org/10.1051/0004-6361/201322616}.
\adsurl{2014A&A...561A.134Z}.
\end{barticle}
\endbibitem

\bibitem[\protect\citeauthoryear{{Zhang}, {Li}, and {Ning}}{2017}]{Zha2017}
\begin{barticle}
\bauthor{\bsnm{{Zhang}}, \binits{Q.M.}},
\bauthor{\bsnm{{Li}}, \binits{D.}},
\bauthor{\bsnm{{Ning}}, \binits{Z.J.}}:
\byear{2017},
\batitle{{Simultaneous Transverse and Longitudinal Oscillations in a Quiescent
  Prominence Triggered by a Coronal Jet}}.
\bjtitle{\apj}
\bvolume{851},
\bfpage{47}.
\doiurl{https://doi.org/10.3847/1538-4357/aa9898}.
\adsurl{2017ApJ...851...47Z}.
\end{barticle}
\endbibitem

\bibitem[\protect\citeauthoryear{{Zhang} et~al.}{2012}]{Zha2012}
\begin{barticle}
\bauthor{\bsnm{{Zhang}}, \binits{Q.M.}},
\bauthor{\bsnm{{Chen}}, \binits{P.F.}},
\bauthor{\bsnm{{Xia}}, \binits{C.}},
\bauthor{\bsnm{{Keppens}}, \binits{R.}}:
\byear{2012},
\batitle{{Observations and simulations of longitudinal oscillations of an
  active region prominence}}.
\bjtitle{\aap}
\bvolume{542},
\bfpage{A52}.
\doiurl{https://doi.org/10.1051/0004-6361/201218786}.
\adsurl{2012A&A...542A..52Z}.
\end{barticle}
\endbibitem

\bibitem[\protect\citeauthoryear{{Zhang} et~al.}{2015}]{Zha2015}
\begin{barticle}
\bauthor{\bsnm{{Zhang}}, \binits{Q.M.}},
\bauthor{\bsnm{{Ning}}, \binits{Z.J.}},
\bauthor{\bsnm{{Guo}}, \binits{Y.}},
\bauthor{\bsnm{{Zhou}}, \binits{T.H.}},
\bauthor{\bsnm{{Cheng}}, \binits{X.}},
\bauthor{\bsnm{{Ji}}, \binits{H.S.}},
\bauthor{\bsnm{{Feng}}, \binits{L.}},
\bauthor{\bsnm{{Wiegelmann}}, \binits{T.}}:
\byear{2015},
\batitle{{Multiwavelength Observations of a Partially Eruptive Filament on 2011
  September 8}}.
\bjtitle{\apj}
\bvolume{805},
\bfpage{4}.
\doiurl{https://doi.org/10.1088/0004-637X/805/1/4}.
\adsurl{2015ApJ...805....4Z}.
\end{barticle}
\endbibitem

\bibitem[\protect\citeauthoryear{{Zhang} et~al.}{2022a}]{Zha2022a}
\begin{barticle}
\bauthor{\bsnm{{Zhang}}, \binits{Q.M.}},
\bauthor{\bsnm{{Chen}}, \binits{J.L.}},
\bauthor{\bsnm{{Li}}, \binits{S.T.}},
\bauthor{\bsnm{{Lu}}, \binits{L.}},
\bauthor{\bsnm{{Li}}, \binits{D.}}:
\byear{2022}a,
\batitle{{Transverse Coronal-Loop Oscillations Induced by the Non-radial
  Eruption of a Magnetic Flux Rope}}.
\bjtitle{\solphys}
\bvolume{297},
\bfpage{18}.
\doiurl{https://doi.org/10.1007/s11207-022-01952-3}.
\adsurl{2022SoPh..297...18Z}.
\end{barticle}
\endbibitem

\bibitem[\protect\citeauthoryear{{Zhang} et~al.}{2022b}]{Zha2022b}
\begin{barticle}
\bauthor{\bsnm{{Zhang}}, \binits{Y.}},
\bauthor{\bsnm{{Zhang}}, \binits{Q.}},
\bauthor{\bsnm{{Song}}, \binits{D.}},
\bauthor{\bsnm{{Li}}, \binits{S.}},
\bauthor{\bsnm{{Dai}}, \binits{J.}},
\bauthor{\bsnm{{Xu}}, \binits{Z.}},
\bauthor{\bsnm{{Ji}}, \binits{H.}}:
\byear{2022}b,
\batitle{{Statistical Analysis of Circular-ribbon Flares}}.
\bjtitle{\apjs}
\bvolume{260},
\bfpage{19}.
\doiurl{https://doi.org/10.3847/1538-4365/ac5f4c}.
\adsurl{2022ApJS..260...19Z}.
\end{barticle}
\endbibitem

\bibitem[\protect\citeauthoryear{{Zheng} et~al.}{2019}]{Zheng2019}
\begin{barticle}
\bauthor{\bsnm{{Zheng}}, \binits{R.}},
\bauthor{\bsnm{{Yang}}, \binits{S.}},
\bauthor{\bsnm{{Rao}}, \binits{C.}},
\bauthor{\bsnm{{Liu}}, \binits{Y.}},
\bauthor{\bsnm{{Zhong}}, \binits{L.}},
\bauthor{\bsnm{{Wang}}, \binits{B.}},
\bauthor{\bsnm{{Song}}, \binits{H.}},
\bauthor{\bsnm{{Li}}, \binits{Z.}},
\bauthor{\bsnm{{Chen}}, \binits{Y.}}:
\byear{2019},
\batitle{{A Confined Partial Eruption of Double-decker Filaments}}.
\bjtitle{\apj}
\bvolume{875},
\bfpage{71}.
\doiurl{https://doi.org/10.3847/1538-4357/ab0f3f}.
\adsurl{2019ApJ...875...71Z}.
\end{barticle}
\endbibitem

\bibitem[\protect\citeauthoryear{{Zhou} et~al.}{2020}]{Zhou2020}
\begin{barticle}
\bauthor{\bsnm{{Zhou}}, \binits{Y.H.}},
\bauthor{\bsnm{{Chen}}, \binits{P.F.}},
\bauthor{\bsnm{{Hong}}, \binits{J.}},
\bauthor{\bsnm{{Fang}}, \binits{C.}}:
\byear{2020},
\batitle{{Simulations of solar filament fine structures and their
  counterstreaming flows}}.
\bjtitle{Nature Astronomy}
\bvolume{4},
\bfpage{994}.
\doiurl{https://doi.org/10.1038/s41550-020-1094-3}.
\adsurl{2020NatAs...4..994Z}.
\end{barticle}
\endbibitem

\end{thebibliography}

\end{article}
\end{document}